\documentclass[preprint,12pt]{elsarticle}

\usepackage{amssymb}
\usepackage{amsmath}

\usepackage{booktabs}

\usepackage{lineno}

\usepackage{siunitx}
\usepackage{tabularx}

\journal{}

\begin{document}

\begin{frontmatter}



\title{PID-like active control strategy for electroacoustic resonators to design tunable single-degree-of-freedom sound absorbers}



\author[mymainaddress]{Xinxin Guo\corref{mycorrespondingauthor}}
\cortext[mycorrespondingauthor]{Corresponding author}
\ead{xinxin.guo@epfl.ch}

\author[mymainaddress]{Maxime Volery}
\author[mymainaddress]{Hervé Lissek}

\address[mymainaddress]{Signal Processing Laboratory LTS2, Ecole Polytechnique F\'ed\'erale de Lausanne, 1015 Lausanne, Switzerland}

\begin{abstract}
Sound absorption at low frequencies still remains a challenge in both scientific research and engineering practice. Natural porous materials are ineffective in this frequency range, as well as acoustic resonators which present too narrow bandwidth of absorption, thus requiring alternative solutions based on active absorption techniques. In the present work, we propose an active control framework applied on a closed-box loudspeaker to enable the adjustment of the acoustic impedance at the loudspeaker diaphragm. More specifically, based on the proportionality between the pressure inside the enclosure and the axial displacement of the loudspeaker diaphragm at low frequencies, we demonstrate both analytically and experimentally that a PID-like feedback control approach allows tuning independently the compliance, the resistance and the moving mass of the closed-box loudspeaker to implement a prescribed impedance of a single-degree-of-freedom resonator. By considering different control combinations to tailor the resonator characteristics, a perfect absorption (with absorption coefficient equal to $1$) is achievable at the target resonance frequency, while enlarging the effective absorption bandwidth. Moreover, the proposed feedback control strategy shows an excellent control accuracy, especially compared to the feedforward-based control formerly reported in the literature. The mismatches between the performance of experimental prototype and the model, likely to result from the control time delay and the inaccuracy in estimating the loudspeaker parameters, can be compensated directly by tuning the control parameters in the control platform. The active resonators implemented through the reported control scheme can be used to build more complex acoustic devices/structures to enable high-efficiency broadband sound absorption or other types of acoustic phenomena such as wavefront shaping.

\end{abstract}




\end{frontmatter}


\section{Introduction} \label{intro}
The research on effective means for reducing low-frequency noise triggers an important interest, owing to their significant impact on human health and everyday activities. The prevailing solution for sound absorption relies on the use of natural sound absorbing materials, such as porous and fibrous media \cite{Biot_1956, Allard_2009} and acoustic resonators \cite{lagarrigue_bamboo_2013}. However, at low frequencies, these materials either may only be effective at the price of an extensive bulkiness comparable to the operating wavelength or provide too narrow bandwidth of absorption, which severely hinders their applications in real life. To overcome such issue, active sound absorbers have received a surge of interest, since they can be constructed in limited space while providing tunable acoustic properties. Moreover, it is well known that for any passive, linear and time-invariant system, the bandwidth and the absorption efficiency are mutually constrained \cite{Bode_1945, Pozar_1998, absorp_constraint_2009}. Active treatments can violate the passivity of materials, thereby leading to efficient sound absorption in a wider frequency range.

For designing active sound absorbers, Electroacoustic Resonators (ERs) \cite{Cox_2016} consisting of loudspeaker systems and behaving as membrane-type absorbers are favored, the acoustic impedance of which can be controlled through connection with external loads. The concept of sound absorption by active electroacoustic means has been formally introduced by Olson and May \cite{Olson_JASA_1953}, who applied a feedback control on an electrodynamic loudspeaker, based on the sensing of sound pressure in the vicinity of the loudspeaker, allowing modification of the acoustic impedance at its diaphragm. Thereafter, various control approaches have been developed on the ERs \cite{control_1997, Active_metasurface_PSheng, boulandet_electrody_2016, Bryk_JSV}, such as shunting the electrical terminals of the loudspeaker with a matched electrical network \cite{etienne_IEEE, boulandet_2010, herve_2011, etienne2016}, or feeding back a current/voltage prescribed from the measured acoustic pressure/velocity \cite{control_1997, ER_control_JASA2003, etienne_IEEE, collet_active_2009, boulandet_2010, herve_2011,  electrody_actuator_JSV_2017, etienne2016}. The latter control method through current/voltage assignment presents more advantages since its digital implementation offers more flexibility in control law definition and makes the reconfiguration of control much easier.

In the most recent concept of Active ER (AER) of Ref. \cite{etienne2016}, the sensed sound pressure is converted into a current driving the ERs in real time, bypassing the electrical impedance, thus avoiding the inherent instabilities owing to the electrical inductance of the coil. The apparent acoustic impedance of the AER can be adjusted accordingly following the desired frequency dependence, leading to a family of designs of Single-Degree-Of-Freedom (SDOF) and Multiple-Degrees-Of-Freedom (MDOF) sound absorbers \cite{etienne_IEEE,etienne_MDOF, etienne_acta, etienne2016}. Such tunability is the key in many applications, such as room mode damping \cite{rivet_aes_2016,etienne2016}, wavefront shaping \cite{lissek_JAP_2018} or aircraft engine tonal noise reduction \cite{boulandet_jsv_2018}. Nevertheless, due to the unavoidable time delay in the control execution, as well as the inaccuracy in the parameter estimation required for control law definition, this type of feedforward-based control is always accompanied with a noticeable mismatch of the achieved acoustic impedance around the natural (passive) resonance of the ER, which could even make the controlled ER lost its passivity and produce instability. In the present work, a novel PID-like control approach is proposed and developed as an alternative, based on the sensing of the pressures inside the enclosure and in front of the diaphragm. With such control architecture, it is expected to adjust independently each characteristic of the resonator (moving mass, resistance and compliance), while enabling the acoustic performance of the realized AER to better match the prescribed target.

The paper is organized as follows. Taking the closed-box loudspeaker characterized in section \ref{ER} as the basic prototype, the principle and the methodology of the proposed PID-like active control are presented in section \ref{AER}. Analytical simulations of the proposed control method are first carried out in section \ref{Simu}, yielding a preliminary evaluation of performance and providing a reference for the following experimental study. Thereafter in section \ref{Exp}, with the set up described in \ref{set up}, the proposed control on the ER is achieved and explored in \ref{AER_results}. Different control laws are considered to adjust the compliance, the moving mass and the resistance of the AER either individually or simultaneously. Finally, the PID-like feedback control is further assessed in section \ref{AER_stability} in terms of control accuracy, by comparing the acoustic performance of the achieved AER with the ones obtained with the feedforward control already developed in Ref. \cite{etienne2016}. 

\section{Electroacoustic resonators} \label{ER}
The electroacoustic resonator (ER) concept refers to membrane-type resonators achieved for example with electrodynamic loudspeakers. At low frequencies and under weak excitation (linear assumption), an electrodynamic loudspeaker behaves as a linear Single-Degree-Of-Freedom (SDOF) ER. Its mechanical part can be modeled as a conventional mass-spring-damper system, in which the moving diaphragm of mass $M_{ms}$ is attached through an elastic suspension of mechanical compliance $C_{ms}$, and the global losses are accounted for in the mechanical resistance $R_{ms}$. While, when a current circulates in the moving coil of the ER, an electromagnetic force is generated and applied simultaneously, actuating the ER diaphragm. Denoting $S_d$ the effective diaphragm surface area and $Bl$ the force factor of the moving coil, the full dynamics of the ER membrane subjected to external acoustic pressures can be described in the time domain as follows:
\begin{equation}
M_{ms} \frac{d v(t)}{d t} =  S_{d}(p_f(t)-p_b(t))-R_{ms}v(t)-\frac{1}{C_{ms}}\int v(t)dt -Bli(t),
\label{eq: ER}  
\end{equation}
where $p_f(t)$ and $p_b (t)$ represent the acoustic pressures applied respectively to the front and the rear faces of the ER membrane, whereas $v(t)$ and $i(t)$ designate the axial inward velocity of the membrane and the electrical current circulating in the moving coil, respectively.

When the loudspeaker is closed with an enclosure, the sound pressure $p_b(t)$ inside the cavity of volume $V_b$ can be assumed uniform at low frequencies (when wavelengths are much larger than the enclosure dimensions), yielding a linear relation with the axial displacement of the loudspeaker diaphragm $\xi(t)=\int v(t)dt$, namely
\begin{equation}
p_b(t) \cong \frac{S_d}{C_{ab}}\xi(t),
\label{eq: pb}
\end{equation}
with $C_{ab}=V_b/(\rho c^2)$ representing the acoustic compliance of the enclosure, where $\rho$ and $c$ denote the air mass density and the associated speed of sound. 

The ER considered in this paper corresponds to such type of closed-box loudspeaker. Following Eq.~\eqref{eq: pb}, the pressure applied to the rear face of loudspeaker membrane $p_b$ can accordingly be substituted by the term related to the displacement $\xi$. The compressibility of the fluid in the enclosure, identified by the introduced compliance $C_{ab}$, can then be accounted for in an overall mechanical compliance, expressed as $C_{mc}=C_{ms}C_{ab}/(S_d^2C_{ms}+C_{ab})$, leading finally to the motion equation of displacement $\xi(t)$ for the ER membrane as follows:
\begin{equation}
M_{ms} \frac{d^2 \xi(t)}{dt^2} =p_f(t)S_{d}-R_{ms}\frac{d \xi(t)}{d t}-\frac{1}{C_{mc}}\xi(t)-Bli(t).
\label{eq: closed-box ER}
\end{equation}

Using the Fourier transform, the acoustic response of the ER membrane can be characterized in the frequency domain by the specific acoustic impedance $Z_s(j\omega)$. For any linear and time-invariant system, it is defined as the transfer function between the acoustic pressure applied to the system and the resulting normal acoustic velocity.

Regarding the typical passive case with open circuit, i.e., with no current circulating in the moving coil ($i(t)=0$), the specific acoustic impedance of the ER takes the form of 
\begin{equation}
Z_{so}(j\omega)=\frac{P_f(j\omega)}{V(j\omega)}=j\omega \frac{M_{ms}}{S_d} +\frac{R_{ms}}{S_d}+\frac{1}{j\omega C_{mc}S_d},
\label{eq: Z_OC}
\end{equation}
where the uppercase symbols $P_f$ and $V$ are used to represent the frequency responses of the considered acoustic quantities (front pressure and normal velocity) to distinguish with their denotations in the time domain (designated by lowercase symbols $p_f$ and $v$ respectively).

The natural resonance of the ER is characterized by the resonance frequency and the quality factor, for the open-circuit case they are defined respectively as
\begin{equation}
f_{so}=\frac{1}{2\pi}\sqrt{\frac{1}{M_{ms}C_{mc}}},\,\,\, Q_{so}=\frac{1}{R_{ms}}\sqrt{\frac{M_{ms}}{C_{mc}}}.
\label{eq: fs Qs}
\end{equation}

Another common passive case is the short circuit configuration, i.e., the two electrical terminals of the loudspeaker are externally connected. In this case, the specific acoustic impedance reads
\begin{equation}
Z_{ss}(j\omega)=j\omega \frac{M_{ms}}{S_d}+\frac{R_{ms}}{S_d}+\frac{1}{j\omega C_{mc}S_d}+\frac{(Bl)^2}{S_d(j\omega L_e +R_e)},
\label{eq: Z_SC}
\end{equation}
where $R_e$ and $L_e$ are the DC resistance and the inductance of the moving coil, respectively. In the low-frequency range typically below $\SI{500}{Hz}$ (assuming $R_e \approx 6 \Omega$ and $L_e \approx \SI{0.1}{mH}$), the inductance is generally small enough compared to $R_e/\omega$ to be neglected.

Before the implementation of the active impedance controls, both the open-circuit and the short-circuit cases will be considered during the calibration stage presented in section \ref{calibration}, they allow numerical fittings to estimate the relevant Thiele/Small parameters \cite{small1972closed-box} of the resonator required for the control law definition.

\section{Active impedance control on ERs} \label{AER}
In the present work, we consider the active impedance control which allows the adjustment of impedance on the actuated membrane through assignment of a feedback electrical current. The main idea is to properly define the control law to enable the transfer from the sensed acoustic quantities to the output current which will be sent back to the ER. Such type of control on current is performed without the need to model the electrical part of the ER, thus it is more stable compared to other types of voltage-based control \cite{boulandet_electrody_2016, etienne2016}.

\subsection{Feedforward-based impedance control approach} \label{AER_pf}
A feedforward-based control approach based on the sensing of the front pressure $p_f(t)$ to implement AERs with tunable impedance properties has been recently reported in Ref. \cite{etienne2016}, named FF-AER control approach in the following. Denoting $Z_{st}$ the target specific acoustic impedance to be achieved through active control, the transfer function $\Phi(j\omega)$ allowing the impedance modification from $Z_{so}$ to $Z_{st}$ can be derived in the frequency domain from Eq.~\eqref{eq: closed-box ER} as
\begin{equation}
\Phi(j\omega) = \frac{I(j\omega)}{P_f(j\omega)}=\frac{S_d}{Bl}\frac{Z_{st}(j\omega)-Z_{so}(j\omega)}{Z_{st}(j\omega)},
\label{eq: Phi}
\end{equation}
where $I(j\omega)$ denotes the Fourier transform of the current $i(t)$.

Focusing on the realization of a SDOF resonator, the target impedance $Z_{st}$ is considered to take a form similar to the passive one of Eq.~\eqref{eq: Z_OC}, namely
\begin{equation}
Z_{st}(j\omega)=j\omega \mu_{M} \frac{M_{ms}}{S_d}+\mu_{R}\frac{R_{ms}}{S_d}+\frac{\mu_{C}}{j\omega C_{mc}S_d},
\label{eq: Zst_pf}
\end{equation}
where $\mu_{M}$, $\mu_{C}$ and $\mu_{R}$ are three design parameters, enabling respectively the moving mass, the compliance and the resistance of the ER to be tuned independently through the control.

Once the impedance control is applied to the ER as prescribed by the law of Eq.~\eqref{eq: Zst_pf}, the resonance of the achieved AER is expected to be adjusted to the frequency
\begin{equation}
f_{st}=\frac{1}{2\pi}\sqrt{\frac{\mu_C}{\mu_M M_{ms}C_{mc}}}=\sqrt{\frac{\mu_C}{\mu_M}}f_{so},
\label{eq: fst_pf}
\end{equation}
and the quality factor is supposed to change into $Q_{st}$ of form
\begin{equation}
Q_{st}=\frac{1}{\mu_R R_{ms}}\sqrt{\frac{\mu_M\mu_CM_{ms}}{C_{mc}}}=\frac{\sqrt{\mu_M\mu_C}}{\mu_R}Q_{so}.
\label{eq: Qst_pf}
\end{equation}

Therefore, the target resonance frequency $f_{st}$ (with respect to that of the natural one $f_{so}$) depends on the ratio $\mu_C/\mu_M$, whereas the ratio $\sqrt{\mu_M\mu_C}/\mu_R$ dominates the resonance bandwidth of the AER. Taking $\mu_R=\mu_C=1$ as an example, when a value of $0.5$ is assigned to the control parameter $\mu_M$, the resonance of the AER should be shifted to the frequency $\sqrt{2} f_{so}$ while its bandwidth increases by a factor $\sqrt{2}$ (since $Q_{st}$ decreases by a factor $\sqrt{2}$).  

\subsection{Proposed PID-like feedback control approach} \label{AER_pb}
\begin{figure}
	\includegraphics[width=1\textwidth]{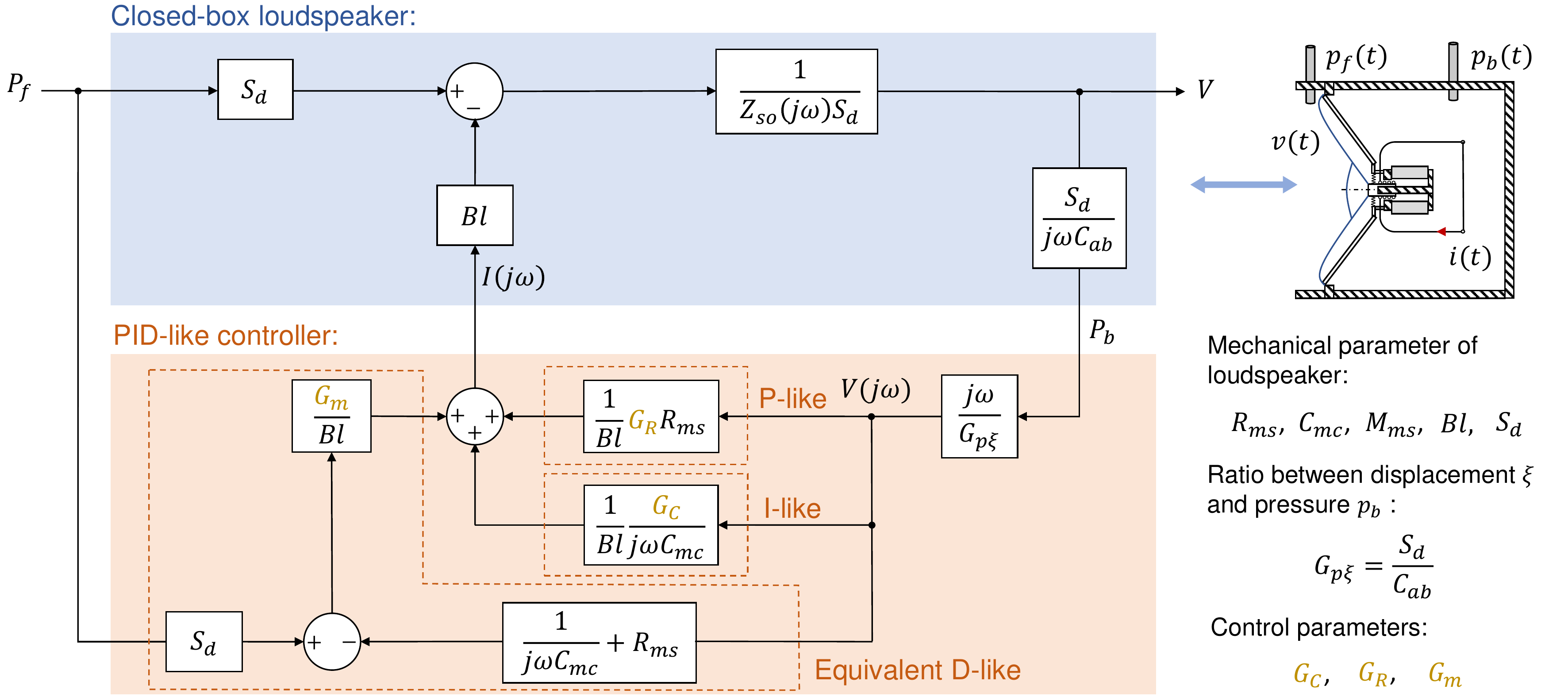}
	\centering
	\caption{\label{fig: control_schema} Block diagram of the closed-box loudspeaker under the proposed PID-like impedance control, based on the sensing of the acoustic pressures in front of the diaphragm $p_f$ and inside the enclosure $p_b$ ($P_f$ and $P_b$ are their frequency responses). $G_C$, $G_R$ and $G_m$ are the three control parameters allowing the compliance, the resistance and the moving mass of the AER to be tuned, respectively. $G_{p\xi}$ denotes the ratio between the pressure $p_b$ and the axial displacement of the AER membrane $\xi$, it is equal to the $(S_d/C_{ab})$ term of Eq.~\eqref{eq: pb}.}
\end{figure}

As an alternative to the FF-AER control approach, the currently proposed feedback control aims at tuning precisely and independently the moving mass, the resistance and the compliance of the resonator in a PID-like spirit. It means, simultaneously applying three individual feedback gains on membrane velocity ("Proportional") to adjust the target resistance, on membrane displacement ("Integral") to adjust the compliance, and on membrane acceleration ("Derivative") to adjust the moving mass. Thus, the proposed method will be referred to as PID-AER in the following. But here, instead of sensing three input signals (through velocimeter, accelerometer, and displacement sensors), it has been decided to simplify the control architecture by relying on only two inputs, namely the front pressure $p_f$ and the pressure inside the AER enclosure $p_b$. Indeed, at low frequencies, the pressure $p_b$ is proportional to the displacement of the membrane as described by Eq.~\eqref{eq: pb}. As a result, the axial velocity of the AER membrane becomes also accessible through the time derivative of the pressure inside the enclosure. The acceleration, instead of being deduced from a second derivative of the displacement estimation which probably introduces instability, it is derived here from an estimation of the overall net forces applied on the membrane, thus requiring the measurement of the front pressure $p_f$. The Fig.~\ref{fig: control_schema} shows the block diaphragm of the AER under such PID-like control. By using two microphones to sense the two pressures $p_f$ and $p_b$, we detail in the following the control strategy and the control laws which enable the compliance, the resistance and the moving mass of ER to be adjusted independently.


\subsubsection{Adjusting the mechanical compliance of the ER (Integral)} \label{AER_pb_xi}
First, we start by tailoring the mechanical compliance of the resonator through assignment of the feedback current with a I-AER method (the I-like control part of Fig.~\ref{fig: control_schema}). Denoting $G_{p\xi}$ the ratio between the pressure inside the enclosure and the normal displacement of the AER membrane ($G_{p\xi}=S_d/C_{ab}$ according to Eq.~\eqref{eq: pb}), we define the current $i(t)$ to be delivered to the loudspeaker terminal as a linear function of $p_b(t)$:
\begin{equation}
i(t) =\frac{G_{C}}{Bl}\frac{1}{C_{mc}}\frac{p_b(t)}{G_{p\xi}}=\frac{G_{C}}{Bl}\frac{1}{C_{mc}}\xi(t),
\label{eq: i(t)_xi}
\end{equation}
with $G_{C}$ the associated gain parameter, to be specified via control platform.

Substituting this expression of $i(t)$ into the motion equation of the AER membrane (Eq.~\eqref{eq: closed-box ER}), we have
\begin{equation}
M_{ms} \frac{d^2 \xi(t)}{dt^2} =p_f(t)S_{d}-R_{ms}\frac{d \xi(t)}{d t}-\frac{1}{C_{mc}}\xi(t)-G_{C}\frac{1}{C_{mc}}\xi(t),
\end{equation}
that is
\begin{equation}
M_{ms} \frac{d^2 \xi(t)}{dt^2} =p_f(t)S_{d}-R_{ms}\frac{d \xi(t)}{d t}-(1+G_{C})\frac{1}{C_{mc}}\xi(t).
\end{equation}
Accordingly, the desired compliance adjustment can be provided by assigning the parameter $G_{C}$. 

\subsubsection{Adjusting the mechanical resistance of the ER (Proportional)} \label{AER_pb_v}
Second, in order to make the resistance of the resonator adjustable, it is mandatory to estimate in real time the normal velocity $v(t)$ of the AER membrane. Based on the sensed pressure inside the enclosure $p_b(t)$ which is related to the membrane displacement, the velocity can be accessed by deriving the pressure $p_b(t)$ in time domain. In the control system, the first-order time derivation is performed with a discrete scheme as 
\begin{equation}
\frac{d p_b}{dt}\approx \frac{p_b(t)-p_b(t-\Delta t)}{\Delta t}
\label{Eq: pb_derivative}
\end{equation}
with $\Delta t$ the chosen step length for the derivative, which could be the inverse of the control sampling frequency or its multiple.

Thereafter, the resulting derivative of $p_b(t)$ can enable the following P-AER type of resistance control law with a linear gain $G_R$ (as shown by the P-like control part of Fig.~\ref{fig: control_schema}):
\begin{equation}
i(t)=\frac{G_R}{Bl} R_{ms}\frac{1}{G_{p\xi}}\frac{d p_b(t)}{dt} =\frac{G_R}{Bl} R_{ms}v(t),
\label{eq: i(t)_v}
\end{equation}
In such a way, the full dynamics of the achieved AER can be re-described with
\begin{equation}
M_{ms} \frac{d^2 \xi(t)}{dt^2} =p_f(t)S_{d}-R_{ms}\frac{d \xi(t)}{d t}-\frac{1}{C_{mc}}\xi(t)-G_R R_{ms}v(t),
\end{equation}
namely
\begin{equation}
M_{ms} \frac{d^2 \xi(t)}{dt^2} =p_f(t)S_{d}-(1+G_R) R_{ms}\frac{d \xi(t)}{d t}-\frac{1}{C_{mc}}\xi(t),
\end{equation}
where the assigned parameter $G_R$ allows the resistance of the resonator to be tuned as specified. Notice that this parameter can only modify the resonance bandwidth and magnitude of the AER, the resonance frequency remains unchanged if the ratio between the compliance and the moving mass is fixed.

\subsubsection{Adjusting the mechanical mass of the ER (Derivative)} \label{AER_pb_a}
The last stage of the control development is dedicated to enabling the adjustment of the moving mass of the resonator through a D-like control approach. To this end, the most straightforward way is applying linearly a gain to the membrane acceleration derived from the second time derivative of the pressure inside the enclosure. However, the introduction of higher order time derivatives will cause a significant increase in response magnitude at high frequencies, which can probably make the system unstable. Moreover, as shown by Eq.~\eqref{Eq: pb_derivative}, the time derivative is carried out in a discrete manner via the control platform, it yields an additional time delay in the control execution. Thus, the adoption of a second derivative can also seriously affect the control accuracy. With the aim of defining a more precise and stable control scheme, we propose herein an alternative method that allows the moving mass of the AER to be adjusted but without need to perform the second derivative, as illustrated by the equivalent D-like control part of Fig.~\ref{fig: control_schema}. In this method, both the pressures inside the enclosure and in front of the AER membrane should be sensed in real time, the feedback current $i(t)$ is defined as follows:
\begin{equation}
i(t)=\frac{G_m}{Bl}\left(p_f(t)S_{d}-R_{ms}\frac{1}{G_{p\xi}}\frac{d p_b(t)}{d t}-\frac{1}{C_{mc}}\frac{1}{G_{p\xi}}p_b(t)\right),
\label{eq: i(t)_a}
\end{equation}
where $G_m$ is the control parameter assigned for adjusting the moving mass.

Then, the substitution of such expression of $i(t)$ into the motion equation of the AER membrane (Eq.~\eqref{eq: closed-box ER}) results in
\begin{equation}
    \begin{split}
M_{ms} \frac{d^2 \xi(t)}{dt^2} =\,\,& p_f(t)S_{d}-R_{ms}\frac{d \xi(t)}{d t}-\frac{1}{C_{mc}}\xi(t) \\
&- G_m\left(p_f(t)S_{d}-R_{ms}\frac{1}{G_{p\xi}}\frac{d p_b(t)}{d t}-\frac{1}{C_{mc}}\frac{1}{G_{p\xi}}p_b(t)\right).      
    \end{split}
\end{equation}

Considering the linear relation between $p_b(t)$ and $\xi(t)$ as described by Eq.~\eqref{eq: pb}, this motion equation can be re-written as:
\begin{equation}
M_{ms} \frac{d^2 \xi(t)}{dt^2} = (1-G_m)p_f(t)S_{d}-(1-G_m)R_{ms}\frac{d \xi(t)}{d t}-(1-G_m)\frac{1}{C_{mc}}\xi(t),  
\end{equation}

namely
\begin{equation}
   \frac{M_{ms}}{(1-G_m)} \frac{d^2 \xi(t)}{dt^2} = p_f(t)S_{d}-R_{ms}\frac{d \xi(t)}{d t}-\frac{1}{C_{mc}}\xi(t),    
\end{equation}
where the assigned parameter $G_m$ must be different from $1$, it enables the moving mass of the AER to be tuned from the natural value $M_{ms}$ to $M_{ms}/(1-G_m)$. 

Therefore, as has been shown in the preceding, based on the sensing of the pressures inside the enclosure and in front of the AER membrane, theoretically it is possible to define different P/I/D-like control laws to allow independently tailoring the compliance, the resistance and the moving mass of the AER. By combining all the associated control laws, the electrical current that makes all the mechanical parameters adjustable takes the form of
\begin{equation}\label{Eq: i(t)_pb}
    \begin{split}
   i(t)=\,\,& \frac{G_{C}}{Bl}\left(\frac{1}{C_{mc}}\frac{p_b(t)}{G_{p\xi}}\right)+\frac{G_R}{Bl}\left( R_{ms}\frac{1}{G_{p\xi}}\frac{d p_b(t)}{dt}\right) \\
 & +\frac{G_m}{Bl}\left(p_f(t)S_{d}-R_{ms}\frac{1}{G_{p\xi}}\frac{d p_b(t)}{d t}-\frac{1}{C_{mc}}\frac{1}{G_{p\xi}}p_b(t)\right).      
    \end{split}
\end{equation}

Accounting for this expression of the feedback current, the motion equation of the membrane of the achieved AER (Eq.~\eqref{eq: closed-box ER}) can be re-formulated as:
\begin{equation}
    \begin{split}
 \frac{M_{ms}}{(1-G_m)}\frac{d^2 \xi(t)}{dt^2} =  p_f(t)S_{d}&-\left(1+\frac{G_R}{1-G_m}\right)R_{ms}\frac{d \xi(t)}{d t}\\
 &-\left(1+\frac{G_C}{1-G_m}\right)\frac{1}{C_{mc}}\xi(t),       
    \end{split}
\end{equation}
leading to the following expression of target impedance  
\begin{equation}
    \begin{split}
Z_{st}(j\omega)=\left(\frac{1}{1-G_m}\right) \frac{j\omega M_{ms}}{S_d}&+\left(1+\frac{G_R}{1-G_m}\right)\frac{R_{ms}}{S_d}\\
&+\left(1+\frac{G_C}{1-G_m}\right)\frac{1}{j\omega C_{mc} S_d}.   
    \end{split}
\end{equation}

When comparing the currently proposed PID-AER control method to the formerly developed FF-AER control presented in the section \ref{AER_pf}, the relation between their control parameters can be identified as:
\begin{equation}
\mu_M=\frac{1}{1-G_m},\,\,\mu_R=1+\frac{G_R}{1-G_m},\,\,\mu_C=1+\frac{G_C}{1-G_m}.   
\end{equation}

With the sake of facilitating the definition of the two control schemes and their comparison, the assigned parameters will be unified in the following to those directly linked to the mechanical characteristics of the passive ER, namely $\mu_M$, $\mu_R$, and $\mu_C$. For these two types of impedance control, one can notice that the definition of the feedback current (Eq.~\eqref{eq: Phi} and Eq.~\eqref{Eq: i(t)_pb}) requires always the pre-estimation of the inherent mechanical parameters of the ER ($M_{ms}$, $C_{mc}$ and $R_{ms}$) as well as the force factor $Bl$. This can be carried out by numerically fitting the experimental outcomes of passive cases, as will be explained in the section \ref{calibration}.

\section{Analytical simulations of the AERs achieved with the proposed PID-like impedance control scheme}\label{Simu}
\subsection{Parameter estimations required for performing the PID-AER impedance control}\label{calibration}
With a view to implementing the proposed impedance control, a preliminary calibration phase is required and presented in this section. As mentioned previously in \ref{AER_pb}, the desired control law of Eq.~\eqref{Eq: i(t)_pb} assumes that the pressure inside the enclosure $p_b$ is proportional to the displacement of the AER membrane $\xi$. Therefore, it is mandatory to ensure in advance the validity of such basic assumption. To this end, the transfer function between $p_b$ and $\xi$ is measured in the frequency domain. The result shown in Fig.~\ref{fig: H(p,u)} confirms experimentally that in the frequency range of interest ($[\SI{50}{Hz},\,\SI{500}{Hz}]$), the pressure $p_b$ presents a linear relation with the displacement of the AER membrane $\xi$. The ratio $G_{p\xi}=P_b/\xi$ required for the definition of the control law is experimentally assessed by averaging the transfer function over the frequency range of interest, and found to be $G_{p\xi} \approx \SI{900}{kPa. m^{-1}}$. 

\begin{figure}
	\includegraphics[width=0.5\textwidth]{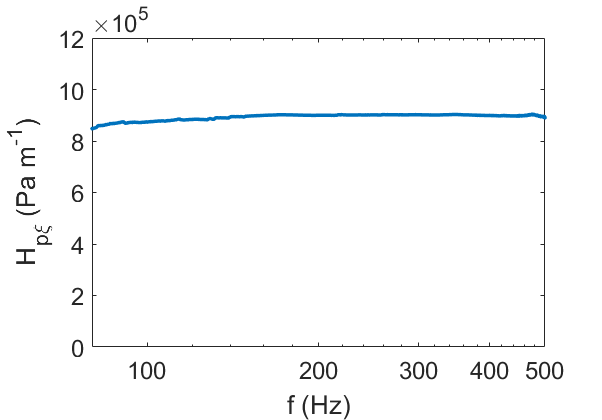}
	\centering
	\caption{\label{fig: H(p,u)} Measured frequency response of the transfer function $H_{p\xi}$ between the pressure inside the enclosure $p_b$ and the axial displacement of the AER membrane $\xi$. }
\end{figure}

Besides $G_{p\xi}$, it is also necessary to estimate the Thiele/Small parameters \cite{small1972closed-box} of the closed-box loudspeaker in use, since their actual values don't often match the ones provided by the manufacturer. For determining these parameters of the passive ER, two calibration measurements of specific acoustic impedance are taken into account. The first is associated with the ER in open circuit case and the second in short circuit case. Theoretically, the specific acoustic impedances in these two cases are described by Eq.~\eqref{eq: Z_OC} and Eq.~\eqref{eq: Z_SC} respectively, as mentioned in section \ref{ER}. Therefore, provided that the effective diaphragm surface area $S_d$ is known, by fitting numerically the impedance magnitude curve in open circuit case following Eq.~\eqref{eq: Z_OC}, one can determine simultaneously the three mechanical parameters of the ER, namely the moving mass $M_{ms}$, the resistance $R_{ms}$ and the overall compliance $C_{mc}$. Notice that the compliance $C_{ms}$ of the loudspeaker itself (without enclosure), known as one of the conventional Thiele/Small parameters \cite{small1972closed-box}, is not necessary to be further calculated, since the estimation of the overall compliance $C_{mc}$ is already sufficient to define the desired control laws. Then the last parameter, the force factor $Bl$, is derived in the same manner as the mechanical parameters but by considering the short circuit configuration. The impedance magnitude of this case is fitted with Eq.~\eqref{eq: Z_SC} where the mechanical parameters are substituted by the values obtained previously when fitting with Eq.~\eqref{eq: Z_OC}. The electrical resistance $R_e$ and also the membrane surface area $S_d$ are considered to be the same as given by the technical specification of the loudspeaker model.

\begin{table}[htbp]
  \caption{\label{ER_parameters} Estimated characteristic parameters of the closed-box Monacor SPX-30M lousdpeaker.}
  \begin{center}
  \begin{tabularx}{0.8\textwidth}{ c | c | c | c | c | c }
    \hline\hline
    \textbf{Parameter} & \multicolumn{1}{c}{$M_{ms}$} & \multicolumn{1}{c}{$R_{ms}$} & \multicolumn{1}{c}{$C_{mc}$} & \multicolumn{1}{c}{$B\ell$} & \multicolumn{1}{c}{$S_d$}\\
    \hline
   \rule{0pt}{12pt} \textbf{Unit} &  \multicolumn{1}{c}{g} &  \multicolumn{1}{c}{N.s.m$^{-1}$} &  \multicolumn{1}{c}{mm.N$^{-1}$} &  \multicolumn{1}{c}{N.A$^{-1}$} &  \multicolumn{1}{c}{cm$^2$}\\
   \hline
   \rule{0pt}{12pt} \textbf{Value} &  \multicolumn{1}{c}{2.80} &  \multicolumn{1}{c}{0.38} &  \multicolumn{1}{c}{0.22} &  \multicolumn{1}{c}{3.59} &  \multicolumn{1}{c}{32}\\
    \hline\hline
  \end{tabularx}
  \end{center}
\end{table}

The estimated characteristic parameters of the used ER are summarized in the table.~\ref{ER_parameters}. For the following studies, the parameter values obtained in this section will be taken into account for both performing the analytical analysis and implementing the proposed impedance control through experiments.

\subsection{Analytical investigation of the proposed PID-AER control approach}\label{Analytical}
Before the experimental realization of an AER, analytical simulations are first carried out in this section to enable a preliminary assessment of the proposed PID-like control approach, especially through the comparison with the FF-AER control method formerly developed in Ref. \cite{etienne2016}. Since there is an inherent time delay in the control execution, the electrical current fed back at the output of the control system is inevitably delayed from initial time $t=0$. Denoting $\tau$ such control time delay, it can be modeled in the frequency domain by applying a factor of $e^{-j\omega \tau}$ to the Fourier transform of the feedback current. For the FF-AER control mentioned in section \ref{AER_pf}, where the control law is determined by a transfer function $\Phi(j\omega)$ (Eq.~\eqref{eq: Phi}) applied to the front pressure $P_f(j\omega)$, the target impedance $Z_{st}(j\omega)$ to be achieved through control will take the following form when accounting for the time delay $\tau$:
\begin{equation}
   Z_{st}(j\omega)=\frac{P_f(j\omega)}{V(j\omega)}=\frac{1}{1-\Phi(j\omega)e^{-j\omega\tau}}Z_{so}(j\omega),
\label{Eq: Z_pf_tau}
\end{equation}
where $Z_{so}$ corresponds to the specific acoustic impedance of the passive ER in the open circuit case as expressed by Eq.~\eqref{eq: Z_OC}.

In the proposed PID-like impedance control, the feedback current is defined as a linear combination of the pressures inside the enclosure ($p_b$) and in front of the AER membrane ($p_f$) (which are subject to different gains), as described previously in section \ref{AER_pb} with Fig.~\ref{fig: control_schema}. Taking into account the linear relation between the pressure $p_b$ and the axial displacement of the loudspeaker membrane $\xi$, the target impedance $Z_{st}(j\omega)$ allowed by the control with time delay can be deduced as:
\begin{equation}
    \begin{split}
    Z_{st}(j\omega)=\,\,& \frac{1}{1-G_me^{-j\omega\tau}}\left[Z_{so}(j\omega)+\left((G_C-G_m)\frac{1}{j\omega C_{mc}S_d}\right.\right.\\
     &\left.\left.+(G_R-G_m) \frac{R_{ms}}{S_d}\frac{1-e^{-j\omega \Delta t}}{j\omega \Delta t}\right)e^{-j\omega \tau}\right],
\label{Eq: Z_pb_tau}  
    \end{split}
\end{equation}
where $\Delta t$ is the time step used for calculating the derivative of $p_b$ (see Eq.~\eqref{Eq: pb_derivative}).

By setting the mechanical parameters of the resonator to the values estimated in section \ref{calibration} (see table.~\ref{ER_parameters}), the specific acoustic impedance of the achieved AER can be analytically computed for both the FF-AER control and the PID-AER control, with Eq.~\eqref{Eq: Z_pf_tau} and Eq.~\eqref{Eq: Z_pb_tau} respectively. The corresponding sound absorption coefficient can then be deduced as
\begin{equation}
\alpha=1-\left| \dfrac{Z_{st}(j\omega)-Z_c}{Z_{st}(j\omega)+Z_c} \right|^2,
\label{eq: alpha}
\end{equation}
where $Z_c=\rho c$ designates the specific acoustic impedance of the air.

\begin{figure}[htbp]
	\includegraphics[width=0.75\textwidth]{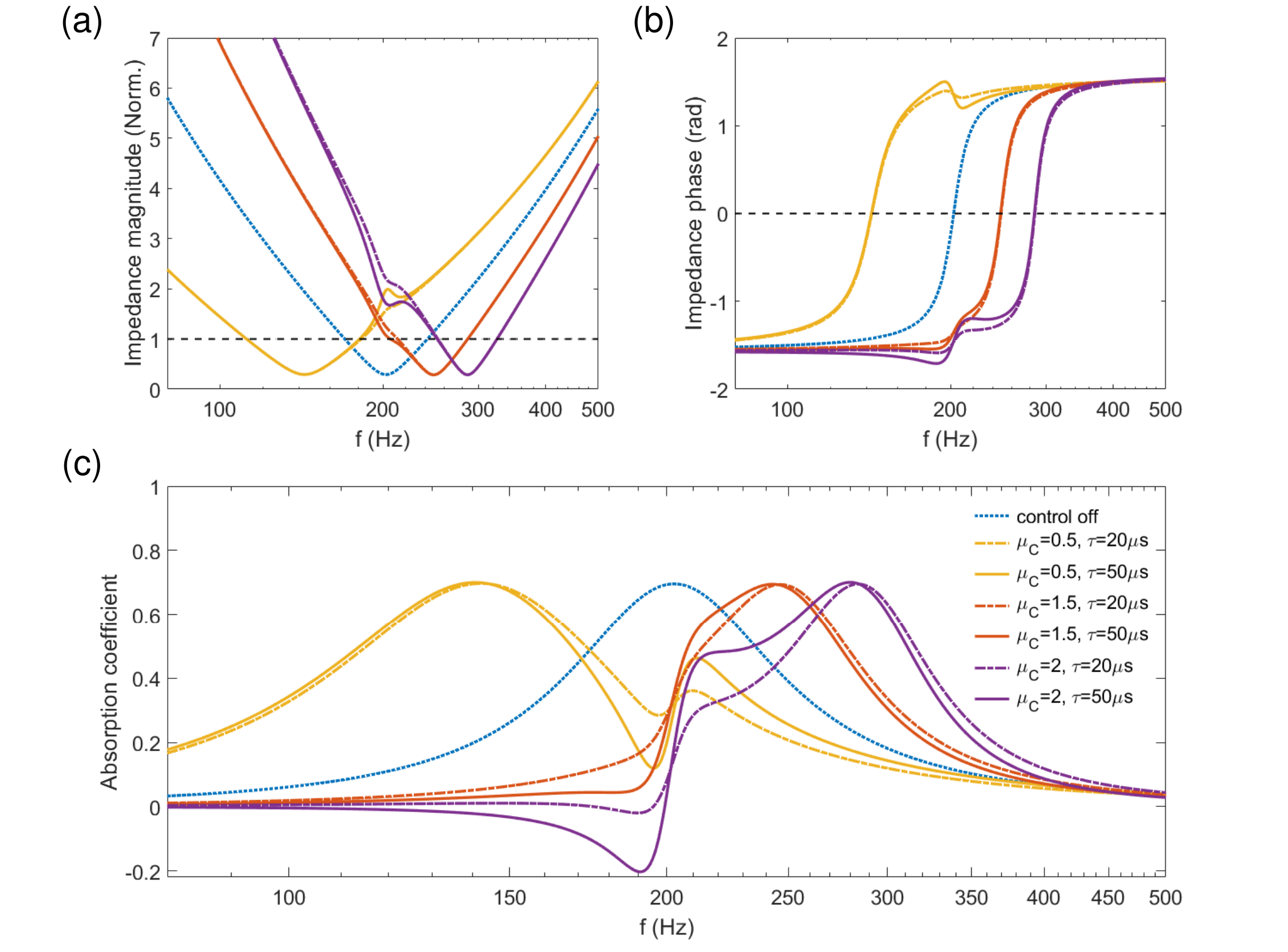}
	\centering
	\caption{\label{fig: Simu_tau_pf} The analytical specific acoustic impedance (magnitude normalized by $Z_c$ (a), phase (b)) and the absorption coefficient (c) of the achieved AER, under the formerly developed FF-AER control \cite{etienne2016} defined with a law of $\mu_C=0.5$ (yellow curves), $1.5$ (red color curves) and $2$ (purple color curves), respectively. The resistance and the moving mass of the controlled ER are preserved with $\mu_R=\mu_M=1$. Two values of control time delay are considered, namely $\SI{20}{\mu s}$ (dash-dotted line curves) and $\SI{50}{\mu s}$ (plain line curves). The legend in (c) is used for all three figures.}
\end{figure}

\begin{figure}[htbp]
	\includegraphics[width=0.75\textwidth]{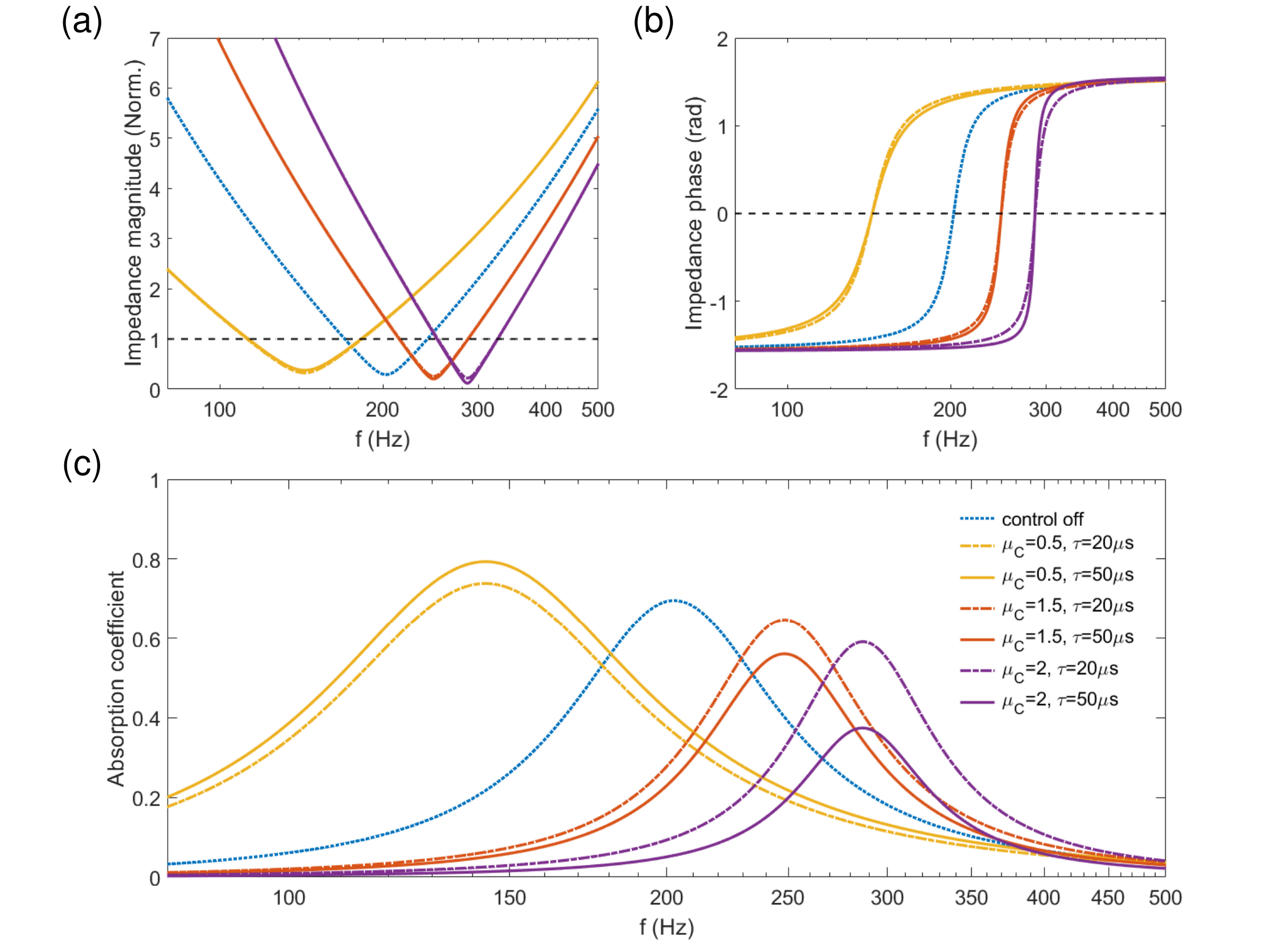}
	\centering
	\caption{\label{fig: Simu_tau_pb} The analytical specific acoustic impedance (magnitude normalized by $Z_c$ (a), phase (b)) and the absorption coefficient (c) of the achieved AER, under the proposed PID-AER control defined with a law of $\mu_C=0.5$ (yellow curves), $1.5$ (red color curves) and $2$ (purple color curves), respectively. The ER resistance and moving mass are preserved with $\mu_R=\mu_M=1$. Two values of control time delay are considered, namely $\SI{20}{\mu s}$ (dash-dotted line curves) and $\SI{50}{\mu s}$ (plain line curves). The legend in (c) is used for all three figures.}
\end{figure}

To better figure out the influence of the time delay on the control accuracy, let us first consider as examples some control cases where only the compliance is adjusted, i.e., with control laws of type $\mu_M=\mu_R=1$ and $\mu_C \neq 1$. The Fig.~\ref{fig: Simu_tau_pf} and Fig.~\ref{fig: Simu_tau_pb} show the control results (specific acoustic impedance (normalized magnitude and phase) and absorption coefficient of the AER) achieved respectively with the FF-AER and the PID-AER control strategies. For each control scheme, two values of time delay are accounted for, namely $\tau=\SI{20}{\mu s}$ (dash-dotted line curves) and $\SI{50}{\mu s}$ (plain line curves), and three compliance control laws are implemented, i.e., $\mu_C=0.5$, $\mu_C=1.5$ and $\mu_C=2$. When the control is executed in an ideal case without time delay, only the resonance frequency of the ER will be shifted (towards low frequency with $\mu_C < 1$ or high frequency with $\mu_C > 1$ according to Eq.~\eqref{eq: fst_pf}). The magnitude and the bandwidth of the impedance and absorption curves are supposed to remain unchanged from the passive configuration (blue dotted line curves in Fig.~\ref{fig: Simu_tau_pf} and Fig.~\ref{fig: Simu_tau_pb}).

Nevertheless, for the FF-AER control strategy, the presence of a time delay can significantly affect its control accuracy, especially around the natural (passive) resonance of the ER, as can be seen on Fig.~\ref{fig: Simu_tau_pf}. It can introduce a remarkable variation on impedance and absorption coefficient, or even make the controlled ER lose its passivity (having negative absorption coefficient) when the time delay is equal to or larger than $\SI{20}{\mu s}$ and when a control law of $\mu_C=2$ is applied (purple color curves of Fig.~\ref{fig: Simu_tau_pf}). Meanwhile, with the proposed PID-AER control scheme which processes the pressure inside the enclosure to tune the compliance (see Eq.~\eqref{Eq: i(t)_pb}), the time delay does not change the profile of the impedance and absorption curves, it mainly yields a marginal effect on the resonance magnitude in the vicinity of the target resonance frequency, as illustrated in Fig.~\ref{fig: Simu_tau_pb}. 

For both the two control methods, one can notice that the larger the time delay, the greater its impact on the resonance behavior. This influence manifests more importantly when the resonance of the AER is adjusted further away from the natural one, as evidenced by the comparison between the control cases of $\mu_C=1.5$ and $\mu_C=2$ (with red and purple color curves respectively) in both Fig.~\ref{fig: Simu_tau_pf} and Fig.~\ref{fig: Simu_tau_pb}.

\begin{figure}[htbp]
	\includegraphics[width=0.75\textwidth]{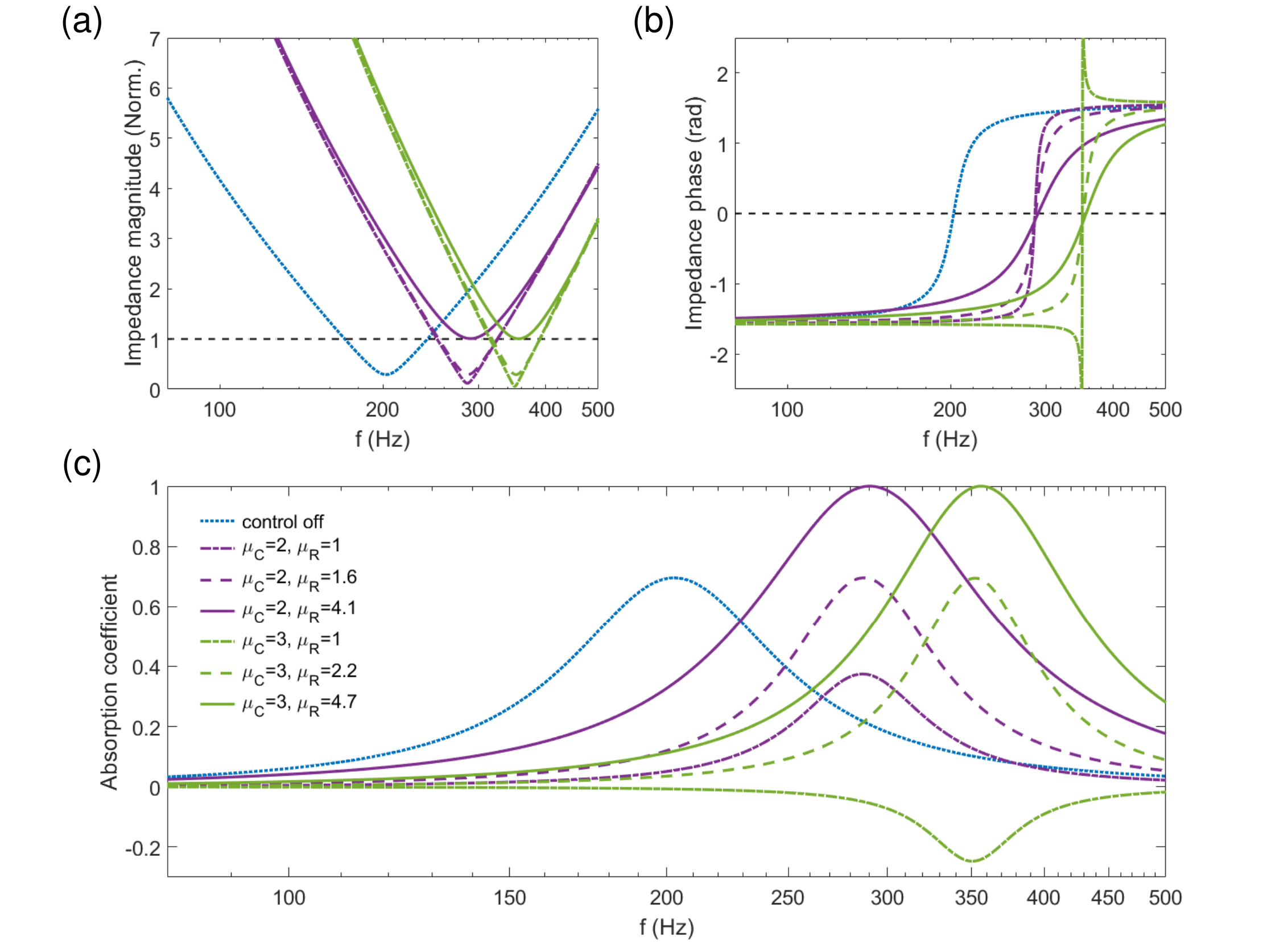}
	\centering
	\caption{\label{fig: Simu_limit} The analytical specific acoustic impedance (magnitude normalized by $Z_c$ (a), phase (b)) and the absorption coefficient (c) of the achieved AER, under the proposed PID-like control defined with a law of $\mu_C=2$ (purple color curves) and $3$ (green color curves), respectively. The control time delay is fixed at $\SI{50}{\mu s}$. For each compliance adjustment with $\mu_C$, three cases are illustrated, namely (i) the resistance of the ER is preserved (dash-dotted line curves), the resistance is increased via $\mu_R$ (ii) to reach the same absorption level as the passive case (dashed line curves) or (iii) to allow a perfect absorption at target resonance frequency (plain line curves). The legend in (c) is used for all three figures. }
\end{figure}

However, with the proposed PID-AER control method, the inaccuracy introduced by the control time delay, although unavoidable, can be compensated by tailoring the target resistance with scaled parameter $\mu_R \neq 1$. For instance, considering the control law of $\mu_C=2$ and a control time delay of $\SI{50}{\mu s}$, the resonance magnitude corresponding to the passive case can be retrieved when the resistance control parameter $\mu_R$ is set to $1.6$, as shown with the purple dashed line curves in Fig.~\ref{fig: Simu_limit}. For the ER excited by plane waves under normal incidence, a perfect absorption is theoretically achieved at resonance for matched acoustic impedance $Z_{st}=Z_c=\rho c$ (with $\mu_R=3.5$). However it requires assigning a larger target resistance with $\mu_R=4.1$ when the compliance is tuned with $\mu_C=2$ and when the same control delay of $\SI{50}{\mu s}$ are accounted for (purple plain line curve of Fig.~\ref{fig: Simu_limit}(c)), which is consistent with the expectations.

Therefore, the proposed PID-AER control strategy appears to be more advantageous compared to the FF-AER method of Ref. \cite{etienne2016} in terms of control accuracy. It allows the impedance and the absorption coefficient of the controlled ER to be adjusted more precisely. The marginal mismatch introduced by the control time delay can be easily compensated by an ad hoc resistance control parameter $\mu_R$. 

Finally, a limit case of control law of $\mu_C=3$ is also studied in this section, as presented in Fig.~\ref{fig: Simu_limit} by green color curves. When the resistance of the ER is simultaneously maintained by setting $\mu_R=1$, a time delay of $\SI{50}{\mu s}$ can result in a negative value of absorption coefficient around the target resonance frequency (green dash-dotted line curves), which means that the controlled ER will inject energy instead of absorbing it. Such situation is likely to lead to an unstable state which should be avoided. Similar to the control case of $\mu_C=2$, this kind of instability can be counteracted by actively increasing the target resistance with $\mu_R >1$. For the case of the compliance adjustment with $\mu_C=3$ allowing the resonance of the AER to be shifted from $f_{so}$ to the target frequency $\sqrt{3}f_{so}$, setting the resistance control parameter to $\mu_R=2.2$ (green dashed line curves) and $\mu_R=4.7$ (green plain line curves) can make the absorption magnitude reach the same level as the passive case and achieve a perfect absorption at the target resonance, respectively.

The analytical study of the current section was aimed at evaluating the feasibility of the proposed PID-AER control strategy and has been limited to a few control settings and cases. In the next section \ref{Exp} where the AER is implemented with an experimental prototype, more control configurations are considered and discussed.

\section{Experimental performance of an AER prototype implemented with the PID-AER control strategy} \label{Exp}

\subsection{Experimental set up}\label{set up}
In our study, the ER prototype is made of a commercially-available Monacor SPX-30M loudspeaker mounted with an enclosure of overall volume $V_b \approx 1 \text{dm}^3$ (with lateral surface of $\SI{12}{cm}\times\SI{12}{cm}$ and with thickness of $\times\SI{6.8}{cm}$). The passive ER is characterized by a natural resonance frequency $f_{so} \approx \SI{200}{Hz}$ in the open circuit case. For a complete characterization of the performance achieved with the various AERs, both the specific acoustic impedance and the sound absorption coefficient are considered, as in the analytical study of section \ref{Analytical}. 
\begin{figure}[htbp]
	\includegraphics[width=0.75\textwidth]{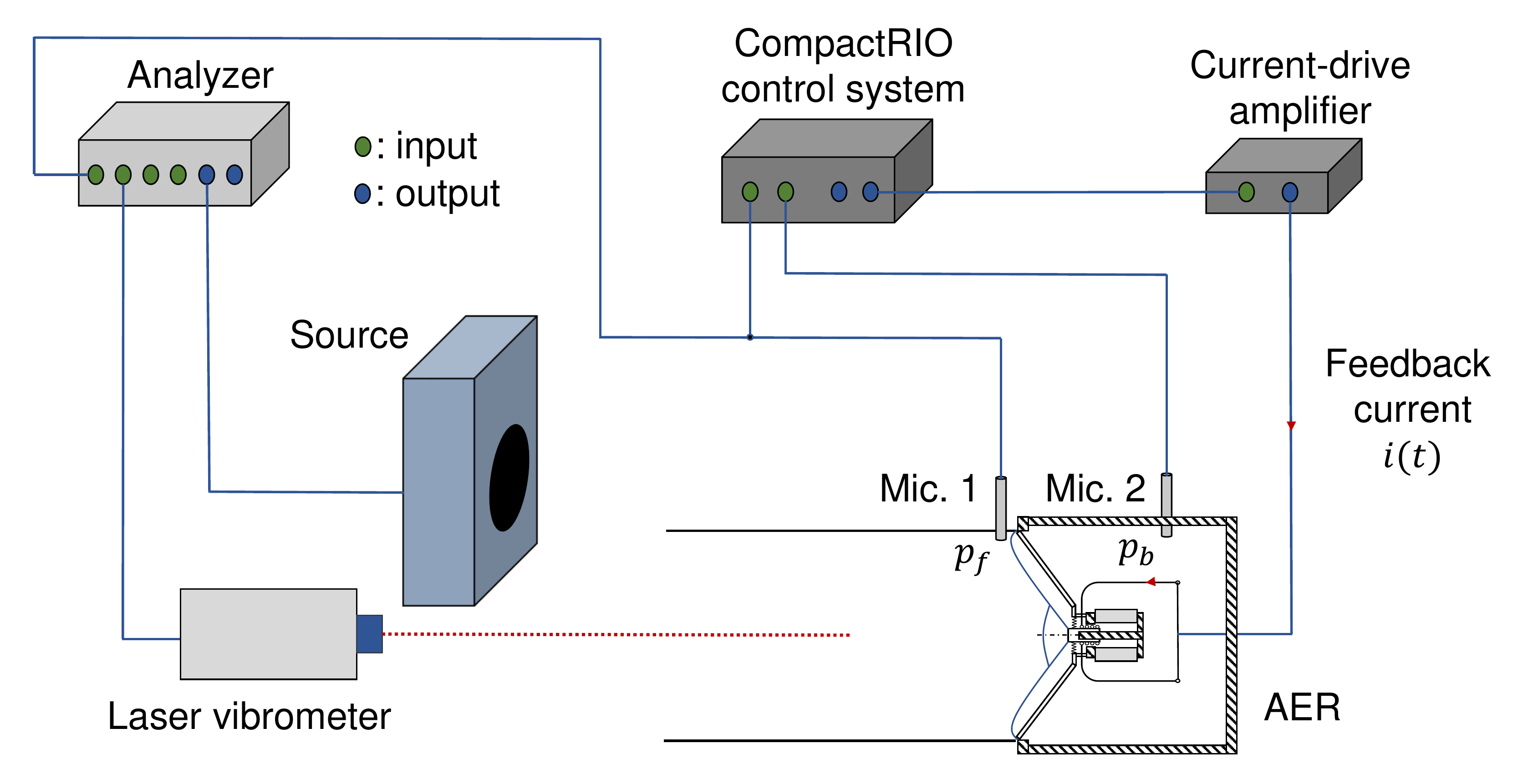}
	\centering
	\caption{\label{fig: set up} Experimental set up. }
\end{figure}

The experimental set up is schematized in Fig.~\ref{fig: set up}. The measurement is carried out with a Bruel \& Kjaer Pulse Multichannel Analyzer that processes the transfer function between multiple input signals, and also handles the output signal generation. A bidirectional sine sweep from $\SI{40}{Hz}$ to $\SI{620}{Hz}$ with sweep rate of $\SI{20}{mdec. s^{-1}}$ is defined and delivered through the Analyzer to drive a Tannoy Reveal Active loudspeaker which is used as the external sound source to excite the controlled ER. The front acoustic pressure $p_f$ and the pressure inside the enclosure $p_b$ are sensed by PCB Piezotronics Type 130D20 ICP microphones (nominal sensitivities $\sigma = \SI{45}{mV.Pa^{-1}}$). The axial velocity of the loudspeaker membrane is captured by a Polytec OFV-500 laser vibrometer and processed with Polytec OFV-5000 controller. Taking the sensed pressures $p_f$ and $p_b$ as the two inputs, a National Instruments CompactRIO system is employed to perform the control on the ER. The control laws are operated on an FPGA platform through LabVIEW 2018 (32bit). The output voltage of the control system is converted into a current through an custom made current-drive amplifier (op-amp based improved Howland current pump circuit \cite{etienne_IEEE}) and then sent back to the electrical terminals of the ER.

With the Multichannel Analyzer, the specific acoustic impedance of the (linear) resonator $Z_s(j\omega)$ can be determined in a straightforward manner, by evaluating the transfer function between the sensed front pressure $p_f$ and the measured membrane axial velocity $v$. Thereafter, the sound absorption coefficient can be derived experimentally by defining a $Z_s$-based function in the Analyzer according to Eq.~\eqref{eq: alpha}. Notice that this definition of sound absorption coefficient is only valid for plane waves under normal incidence. Therefore, we employ a tube of length around $\SI{25}{cm}$ to attach to the front side of the controlled ER in order to guide plane waves towards it.

\subsection{Results of impedance control}\label{AER_results}
Relying on the control system and the experimental set up presented in section \ref{set up}, this section shows the measurement results of specific acoustic impedance and absorption coefficient of the controlled ER. On the FPGA platform, we choose a sampling frequency of $F_e=\SI{40}{kHz}$ for the control implementation, above which its effect on the control accuracy is proved by experiments to be negligible.

\begin{figure}[t]
	\includegraphics[width=0.75\textwidth]{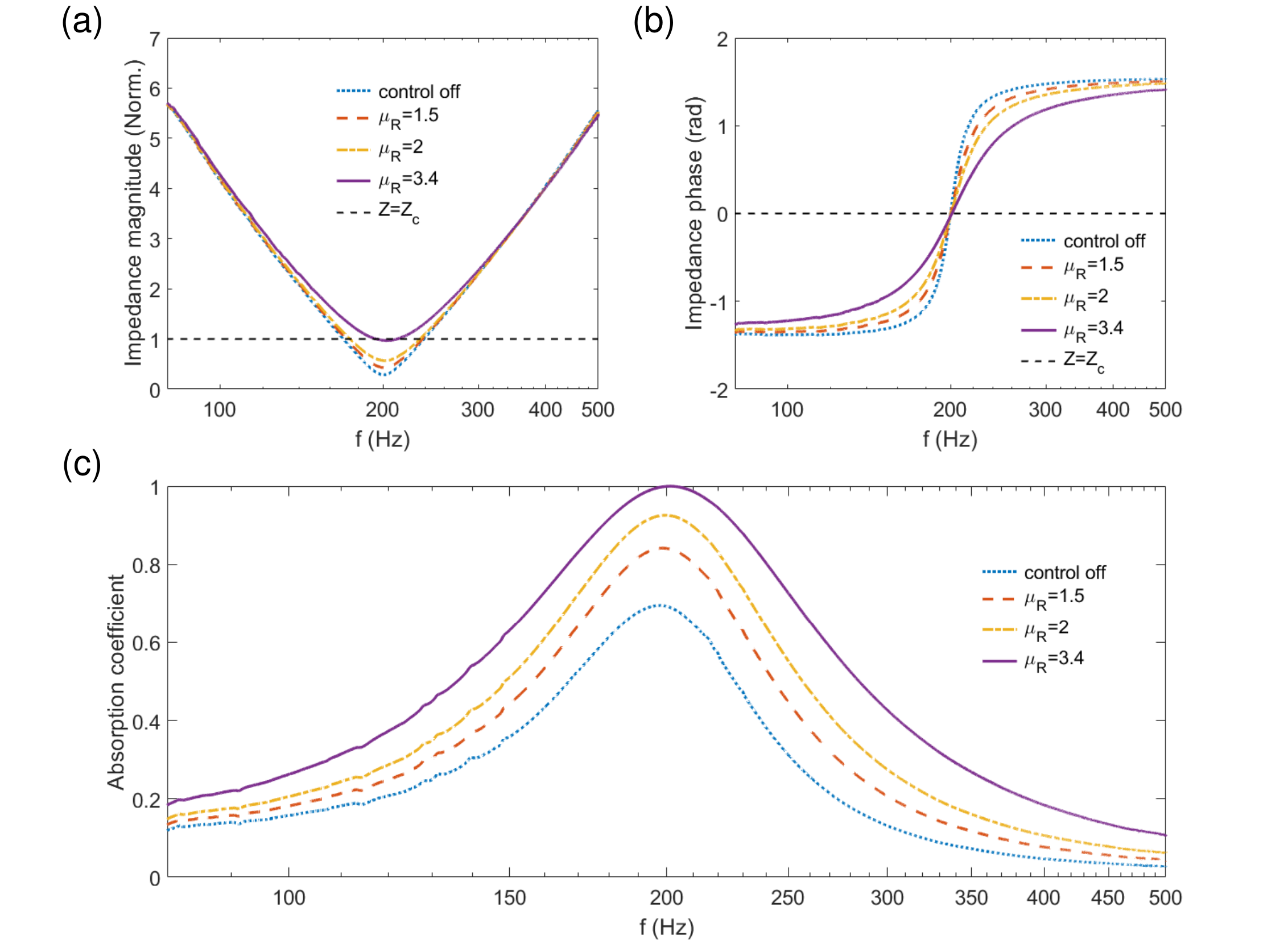}
	\centering
	\caption{\label{fig: Exp_muR} The measured specific acoustic impedance (magnitude normalized by $Z_c$ (a), phase (b)) and the absorption coefficient (c) of the achieved AER, subjected to the proposed PID-AER control with a law of $\mu_R=1.5$ (red dotted line curves), $2$ (yellow dash-dotted line curves) and $3.4$ (purple plain line curves). The compliance and the moving mass of the resonator are maintained by defining $\mu_M=\mu_C=1$. }
\end{figure}

In the passive case, the ER in use is characterised by an absorption coefficient up to around $0.7$ at its natural resonance frequency (around $\SI{200}{Hz}$). The specific acoustic impedance (normalized magnitude and phase) and the absorption coefficient of the passive ER are displayed in Fig.~\ref{fig: Exp_muR} by the blue dotted curves. In order to improve the absorption performance of the ER, we start first by only adjusting its resistance through the proposed PID-AER control. This is achieved by feeding back a current proportional to the axial velocity of ER membrane to the loudspeaker terminals, as explained in the section \ref{AER_pb_v}. In this control scheme, the required membrane velocity is not sensed with the measurement velocimeter but derived in a discrete manner from the sensed pressure inside the enclosure $p_b$ which relates to the membrane displacement, as described by Eq.~\eqref{eq: i(t)_v}. Therefore, the time step $\Delta t$ for calculating the derivative of $p_b$ can probably affect the precision of resistance manipulation. The smallest time step that we can set in the control system corresponds to the inverse of the sampling frequency $1/F_e$. For the considered frequency range of interest ($[\SI{50}{Hz}, \SI{500}{Hz}]$), we confirm experimentally that when varying $\Delta t$ from $1/F_e$ until $4/F_e$ (=$\SI{0.1}{ms}$), no significant impact is observed on the control accuracy. The resulting impedance and absorption curves remain unchanged.

Therefore, for all the following control implementations, the time step $\Delta t$ is fixed at $\SI{0.1}{ms}$ for deriving the axial velocity of the ER membrane. The resistance of the ER is then adjusted through applying a gain to the derived membrane velocity. By setting $\mu_R$ to $1.5$, $2$ and $3.4$ respectively, Fig.~\ref{fig: Exp_muR} shows the corresponding measured specific acoustic impedance (magnitude and phase) and absorption coefficient curves of the achieved AER. Since the compliance and the mass of the passive ER are kept unchanged by imposing $\mu_C=\mu_M=1$, the resonance of the AER occurs at the natural resonance frequency $f_{so}$, as evidenced by Fig.~\ref{fig: Exp_muR}. At this frequency, the passive ER has a purely real impedance with value smaller than that of the impedance of the air $Z_c$. Accordingly, increasing the ER resistance with $\mu_R > 1$ allows the absorption performance of the controlled ER to be improved, as can be noticed by the results in Fig.~\ref{fig: Exp_muR}. When the assigned parameter $\mu_R$ is set to $3.4$ (purple plain line curves), the impedance magnitude of the AER can catch up that of the air $Z_c$ (indicated by the black dashed line curves), thereby leading to a perfect absorption ($\alpha=1$) at its resonance frequency and enabling an effective absorption ($\alpha>0.83$ as detailed in Ref. \cite{etienne_IEEE}) within the range of $[\SI{170}{Hz}, \SI{236}{Hz}]$. In the following, the compliance or/and the moving mass of the AER will be tuned together with the resistance with a view to optimizing its absorption performance.

\begin{figure}[htbp]
	\includegraphics[width=0.75\textwidth]{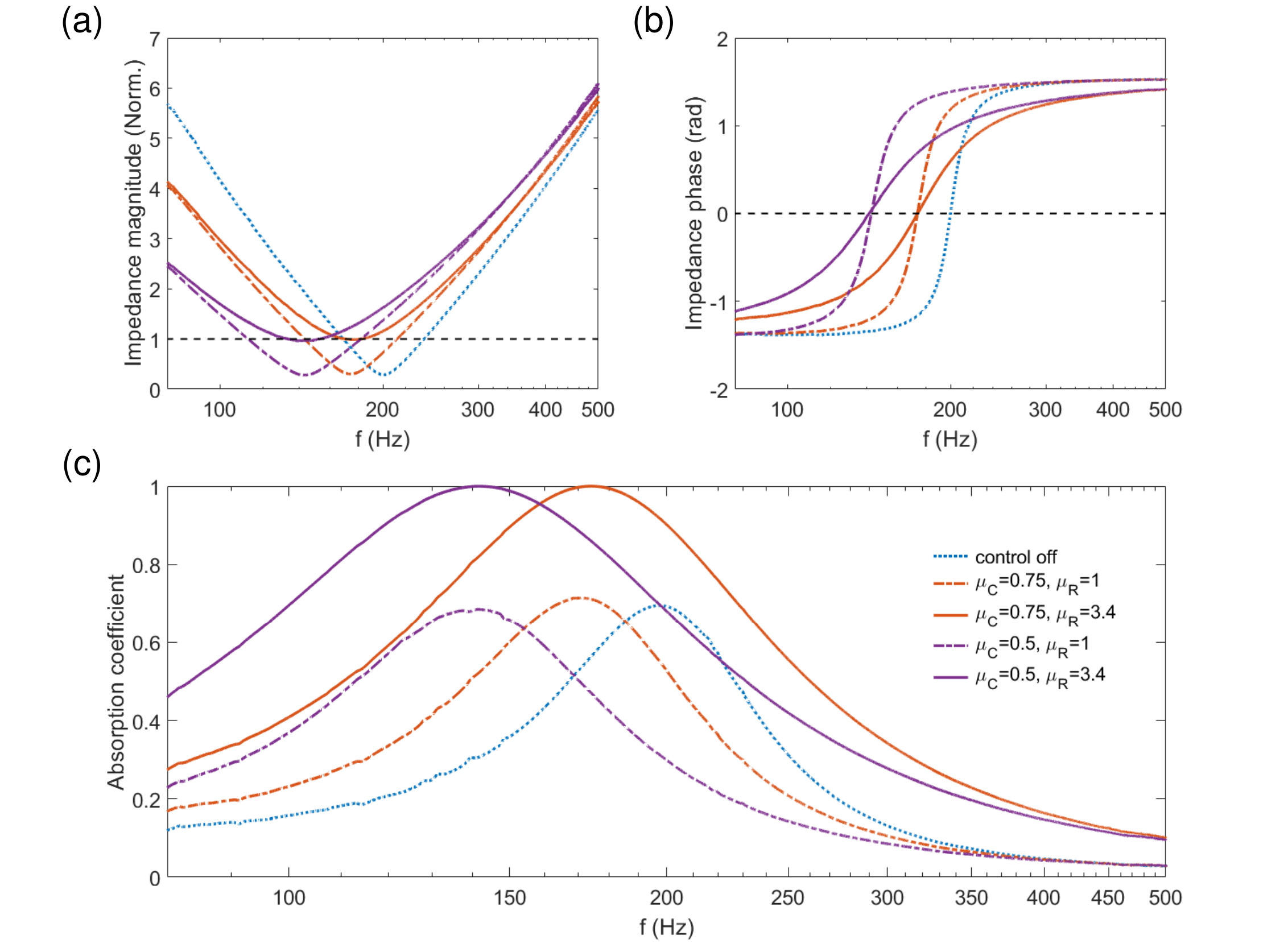}
	\centering
	\caption{\label{fig: Exp_muC_left} The measured specific acoustic impedance (magnitude normalized by $Z_c$ (a), phase (b)) and the absorption coefficient (c) of the achieved AER, subjected to the proposed PID-AER control with a law of $\mu_C=0.75$ and $0.5$. The moving mass of the controlled ER is maintained with $\mu_M=1$. The resistance is tuned to achieve a perfect absorption at the target resonance frequency (plain line curves). The legend in (c) is used for all three figures.}
\end{figure}

\begin{figure}[t]
	\includegraphics[width=0.75\textwidth]{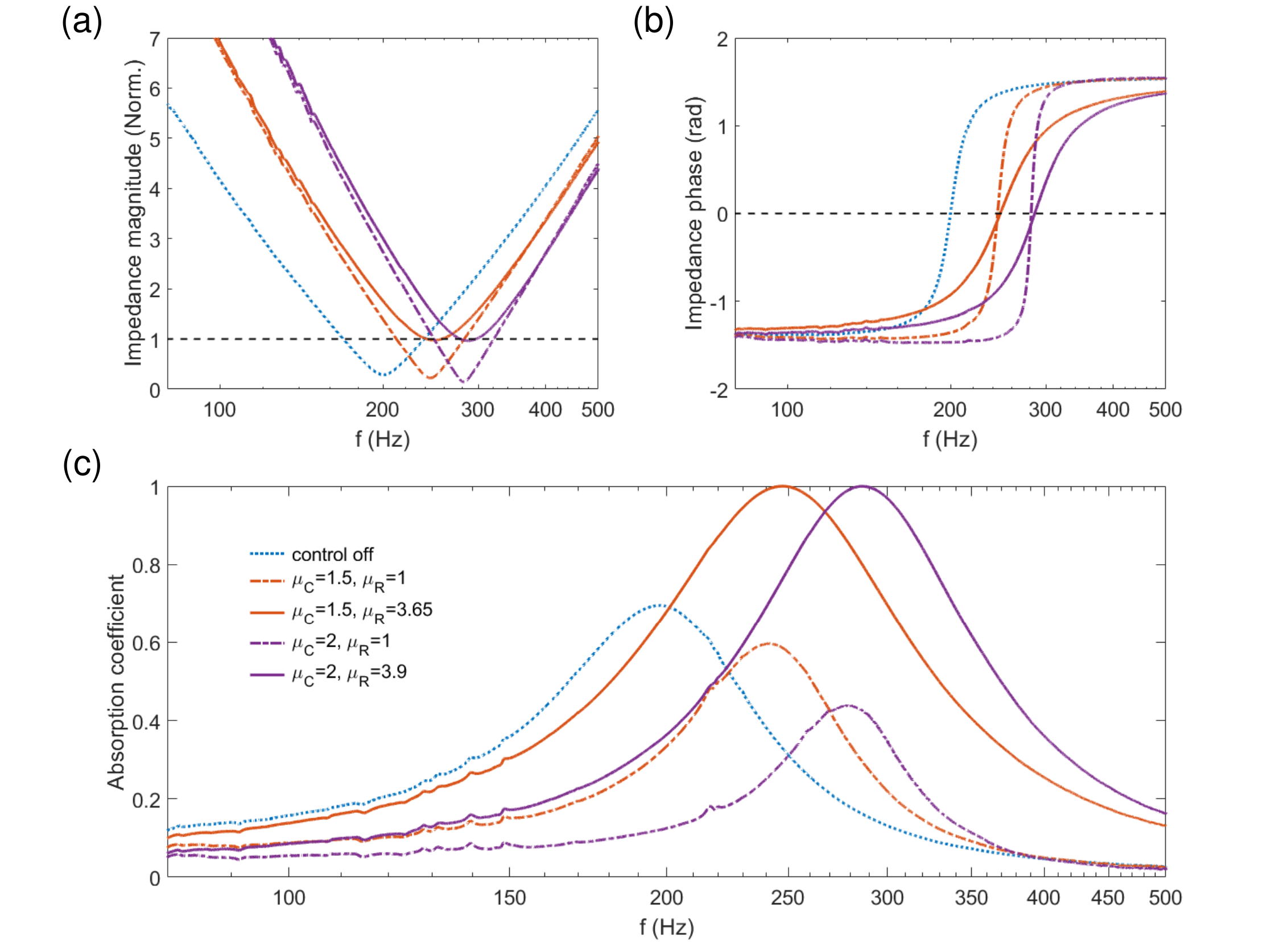}
	\centering
	\caption{\label{fig: Exp_muC_right} The measured specific acoustic impedance (magnitude normalized by $Z_c$ (a), phase (b)) and the absorption coefficient (c) of the achieved AER, subjected to the proposed PID-AER control with a law of $\mu_C=1.5$ and $2$. The moving mass of the controlled ER is maintained with $\mu_M=1$. The resistance is tuned to achieve a perfect absorption at the target resonance frequency (plain line curves). The legend in (c) is used for all three figures.}
\end{figure}

Next, Fig.~\ref{fig: Exp_muC_left} and Fig.~\ref{fig: Exp_muC_right} illustrate the control results when the compliance of the AER is tuned through the control parameter $\mu_C$. These experimental assessments replicate the parameter settings of the analytical study of section \ref{Analytical}, that is, firstly $\mu_C=0.75$ and $0.5$ (Fig.~\ref{fig: Exp_muC_left}), and secondly $\mu_C=1.5$ and $2$ (Fig.~\ref{fig: Exp_muC_right}). Meanwhile, the mass control parameter $\mu_M$ is set to $1$ to preserve the effective moving mass of the resonator. In this case, the resonance of the AER should move from the frequency $f_{so}$ (of open circuit passive case) to the target frequency defined by $f_{st}=\sqrt{\mu_C}f_{so}$. The experimental results presented in Fig.~\ref{fig: Exp_muC_left} and Fig.~\ref{fig: Exp_muC_right} confirm such prediction. For the four chosen values of $\mu_C$ ($0.75$, $0.5$, $1.5$ and $2$), the resonance of the AER is tuned from $\SI{200}{Hz}$ to $\SI{172}{Hz}$ (red color curves in Fig.~\ref{fig: Exp_muC_left}), $\SI{141}{Hz}$ (purple color curves in Fig.~\ref{fig: Exp_muC_left}), $\SI{242}{Hz}$ (red color curves in Fig.~\ref{fig: Exp_muC_right}) and $\SI{280}{Hz}$ (purple color curves in Fig.~\ref{fig: Exp_muC_right}), respectively. Hence, the proposed control approach allows the acoustic properties of the AER to be adjusted as theoretically expected.  

Regarding the control time delay, its influence highlighted in the analytical study of section \ref{Analytical} is also confirmed here by the experimental results. In an ideal control case without any time delay, the control law of type $\mu_C \neq 1$ and $\mu_R=\mu_M=1$ theoretically leads to a resonance frequency shift, while the extreme values of impedance magnitude and absorption coefficient should remain the same as those of the passive case. However, the existence of a time delay in the control execution has been proven to increase these magnitudes at resonance if $\mu_C < 1$ or to decrease them when $\mu_C > 1$, as demonstrated both analytically in section \ref{Analytical} and experimentally herein (dash-dotted line curves of Fig.~\ref{fig: Exp_muC_left} and Fig.~\ref{fig: Exp_muC_right}). Moreover, when a larger compliance adjustment is performed, the magnitude mismatch due to the control time delay becomes more significant, as witnessed by the cases of $\mu_C=1.5$ and $\mu_C=2$ reported on Fig.~\ref{fig: Exp_muC_right}. Comparing the achieved experimental results and the analytical study of section \ref{Analytical}, the actual time delay of the PID-AER control prototype is identified as $\tau_{exp} \approx \SI{50}{\mu s}$.

Based on the aforementioned compliance adjustments, the absorption performance of the AER can be further improved by simultaneously tuning the resistance with $\mu_R$. In Fig.~\ref{fig: Exp_muC_left} and Fig.~\ref{fig: Exp_muC_right}, the plain line curves represent the optimal absorption results achieved when assigning both $\mu_C$ and $\mu_R$. That is, for the four considered compliance control laws defined by $\mu_C=0.75$, $0.5$, $1.5$ and $2$, the absorption coefficient is optimized (reaching $1$ at the target resonance frequency) when the resistance is simultaneously adjusted with $\mu_R=3.4$, $3.4$, $3.65$ and $3.9$ respectively. One can notice that the farther the target resonance (at frequency $\sqrt{\mu_C}f_{so}$) is from the natural one (at frequency $f_{so}$), the larger the required value of $\mu_R$ for achieving perfect absorption. The proposed control strategy provides an opportunity to compensate the discrepancy caused by the control time delay simply through an adjustment of the control parameter $\mu_R$, thereby enabling the absorption performance of the controlled ER to be accurately optimized.

Up to now, the assigned parameter $\mu_M$ has been fixed to $1$ to preserve the effective moving mass of the controlled resonator. Nonetheless, the resonance frequency of the AER depends actually on the ratio $\sqrt{\mu_C/\mu_M}$ (Eq.~\eqref{eq: fst_pf}). Thus acting also on parameter $\mu_M$ should allow spanning a wider range of target resonance frequencies. Then, the pressures both inside the enclosure and in front of the membrane are now used to perform the mass control law of Eq.~\eqref{eq: i(t)_a}. Instead of defining a current directly linked to the inertia of the membrane (e.g., by using an accelerometer on the diaphragm instead), the mass adjustment is enabled here alternatively through applying an overall gain on all the applied forces, as expressed by Eq.~\eqref{eq: i(t)_a} and presented in Fig.~\ref{fig: control_schema}. As a result, the pressure inside the enclosure only needs to be derived once to determine the velocity of AER membrane, its second-order time derivative is avoided, which is crucial for the control stability.

\begin{figure}[t]
	\includegraphics[width=0.75\textwidth]{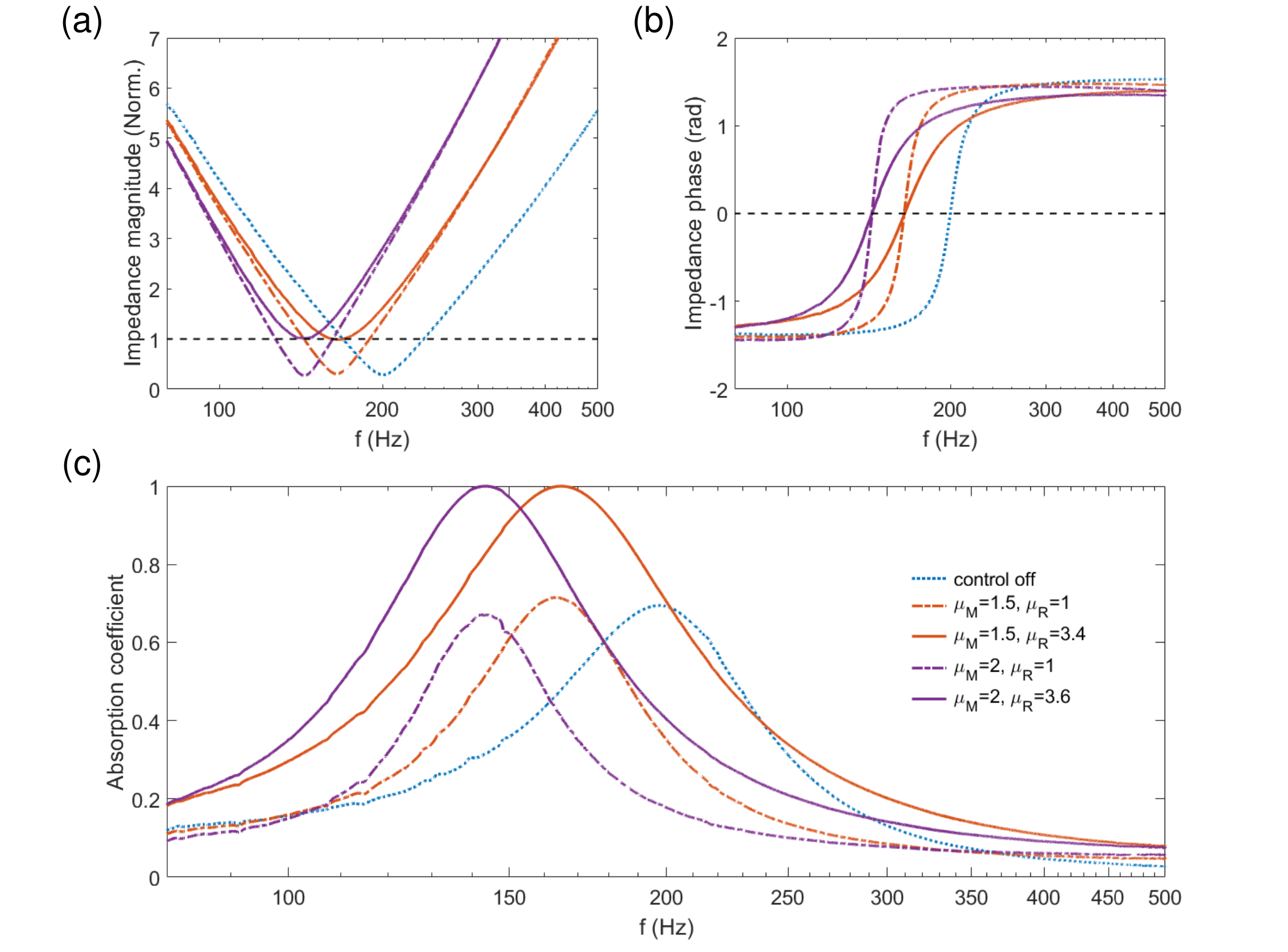}
	\centering
	\caption{\label{fig: Exp_muM_left} The measured specific acoustic impedance (magnitude normalized by $Z_c$ (a), phase (b)) and the absorption coefficient (c) of the achieved AER, subjected to the PID-AER control with a law of $\mu_M=1.5$ and $2$. The compliance of the ER is maintained with $\mu_C=1$. The resistance is tuned to achieve a perfect absorption at the target resonance frequency (plain line curves). The legend in (c) is used for all three figures.}
\end{figure}

\begin{figure}[t]
	\includegraphics[width=0.75\textwidth]{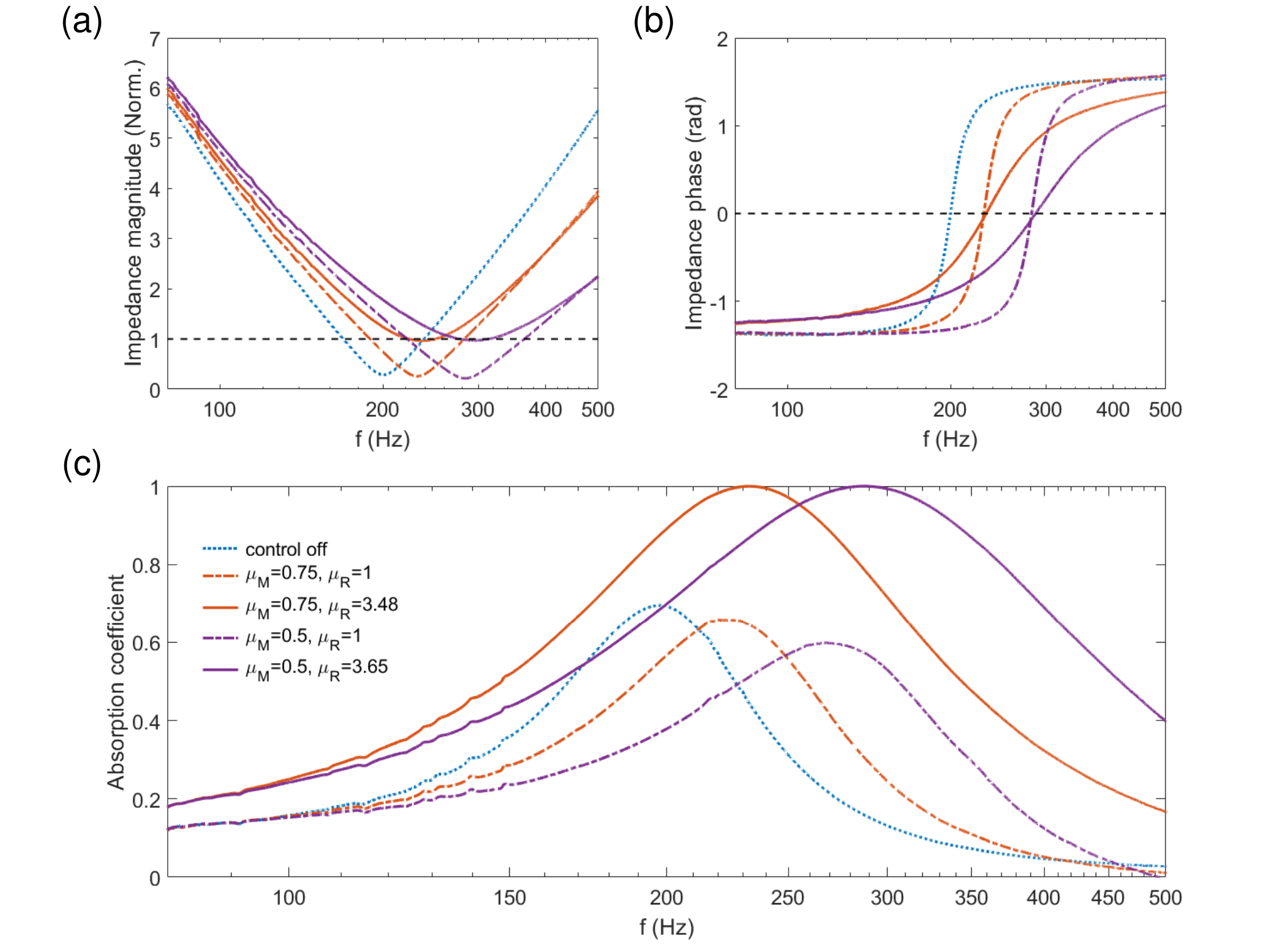}
	\centering
	\caption{\label{fig: Exp_muM_right} The measured specific acoustic impedance (magnitude normalized by $Z_c$ (a), phase (b)) and the absorption coefficient (c) of the achieved AER, subjected to the PID-AER control with a law of $\mu_M=0.75$ and $0.5$. The compliance of the ER is maintained with $\mu_C=1$. The resistance is tuned to achieve a perfect absorption at the target resonance frequency (plain line curves). The legend in (c) is used for all three figures.}
\end{figure}

Similarly to the previously presented controls with $\mu_C$, Fig.~\ref{fig: Exp_muM_left} and Fig.~\ref{fig: Exp_muM_right} show the impedance (magnitude and phase) and the absorption coefficient curves of the AER with only mass control (dash-dotted line curves) and with a simultaneous control of mass and resistance (plain line curves). The assigned parameter $\mu_M$ is set to $1.5$ and $2$ in Fig.~\ref{fig: Exp_muM_left}, and to $0.75$ and $0.5$ in Fig.~\ref{fig: Exp_muM_right}, allowing the resonance frequency of the AER to be tuned to $\SI{164}{Hz}$, $\SI{143}{Hz}$, $\SI{231}{Hz}$ and $\SI{283}{Hz}$, respectively. These results present a good agreement with the analytical simulations. The precision of the proposed mass control method is accordingly confirmed, allowing the resonance of the AER to be fully adjusted as required.

Moreover, the same influence of the control time delay is observed here as in the compliance control cases, which can also be directly counteracted by increasing $\mu_R$. When the resistance is simultaneously tuned, the absorption performance of the controlled ER can be optimized to the greatest extent, making the perfect absorption to be achieved at the target resonance frequency. For the four considered mass control cases ($\mu_M=1.5$, $2$, $0.75$ and $0.5$), perfect absorption ($\alpha = 1$) is achieved at resonance when $\mu_R$ is set to $3.4$, $3.6$, $3.48$ and $3.65$, respectively.

\begin{figure}[htbp]
	\includegraphics[width=0.75\textwidth]{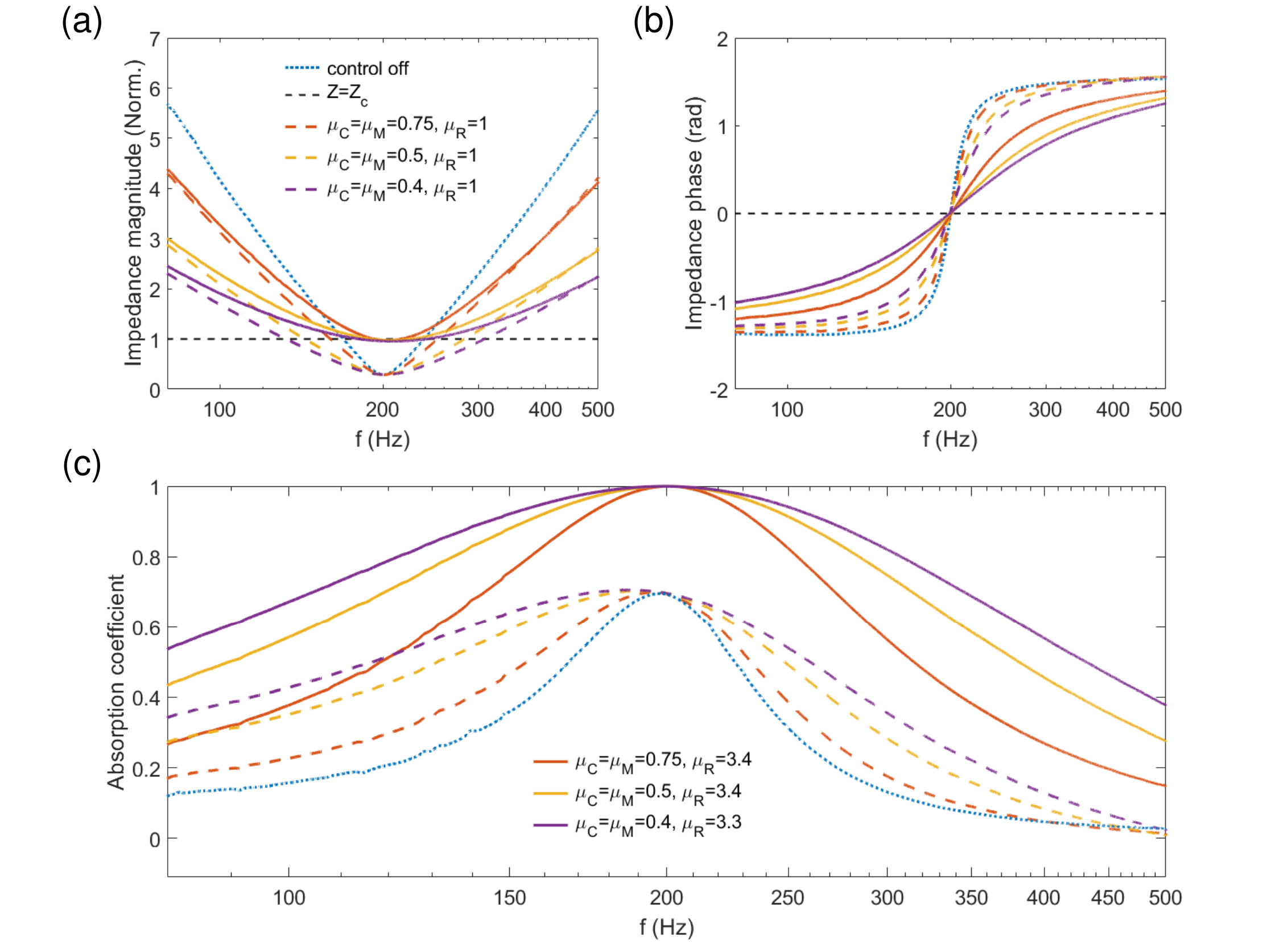}
	\centering
	\caption{\label{fig: Exp_muMC} The measured specific acoustic impedance (magnitude normalized by $Z_c$ (a), phase (b)) and the absorption coefficient (c) of the achieved AER, subjected to the proposed PID-AER control with a law of $\mu_M=\mu_C=0.75$, $0.5$ and $0.4$. The resistance is tuned to achieve a perfect absorption at the target resonance frequency (plain line curves). The legends in (a) and (c) are used for all three figures.}
\end{figure}

After the successive validations of the compliance and the mass adjustments, it is also possible to combine them in order to further improve the absorption performance of the AER. Fig.~\ref{fig: Exp_muMC} illustrates the impedance (magnitude and phase) and the absorption coefficient curves when the control laws of type $\mu_M=\mu_C \neq 1$ are considered. The dashed line curves show the performance of the control cases when the resistance of the ER remains unchanged (with $\mu_R=1$). The red dashed curves represent the control case of $\mu_M=\mu_C=0.75$, the yellow dashed curves and the purple dashed curves show the cases of $\mu_M=\mu_C=0.5$ and $\mu_M=\mu_C=0.4$, respectively. Reminding that the resonance bandwidth of the AER, characterized by the quality factor $Q_{st}$, relies on the term of $\sqrt{\mu_M\mu_C}/\mu_R$ (see Eq.~\eqref{eq: Qst_pf}). Therefore, adjusting simultaneously the compliance and the moving mass with $\mu_M=\mu_C < 1$ leads to an extension of the absorption bandwidth. The smaller the assigned parameters $\mu_C$ and $\mu_M$, the larger the absorption bandwidth of the AER, whereas the resonance frequency is preserved with $\mu_{M}=\mu_C$ (according to Eq.~\eqref{eq: fst_pf}), as can be seen from Fig.~\ref{fig: Exp_muMC}.

The plain line curves in Fig.~\ref{fig: Exp_muMC} show the optimal absorption results achieved by combining the control of resistance with the simultaneous controls of mass and compliance of type $\mu_C=\mu_M \neq 1$. For the control settings of $\mu_M=\mu_C=0.75$, $0.5$ and $0.4$ considered previously, the values of parameter $\mu_R$ yielding a perfect absorption ($\alpha=1$) at resonance frequency are $3.4$, $3.4$ and $3.3$ respectively. Comparing to the passive ER characterized by a maximum absorption coefficient $\alpha$ of around $0.7$, applying the proposed PID-AER control on the ER enables a broader frequency range of effective sound absorption ($\alpha > 0.83$ as detailed in Ref. \cite{etienne_IEEE}). When the control law of $\mu_M=\mu_C=0.4$ and $\mu_R=3.3$ is imposed, the absorption coefficient of the AER is higher than $0.83$ within the frequency range of $[\SI{130}{Hz}, \SI{300}{Hz}]$.

In this section, the PID-AER control method described in \ref{AER_pb} has been experimentally validated. It has been demonstrated experimentally that, based on the sensing of both the pressures inside the enclosure and in front of the AER membrane, one can tailor in a prescribed manner the resonator's effective compliance, resistance and moving mass, independently or simultaneously. In order to evaluate thoroughly the proposed control scheme, we will discuss in the next section \ref{AER_stability} its advantages in terms of control accuracy, especially compared with the FF-AER method developed in Ref.\cite{etienne2016} and described in section \ref{AER_pf}. The stability of the control will also be investigated in \ref{AER_stability} to overview the limitation and efficiency of the method. 
 
\subsection{Assessment of the proposed impedance control approach}\label{AER_stability}
\begin{figure}[htbp]
	\includegraphics[width=0.75\textwidth]{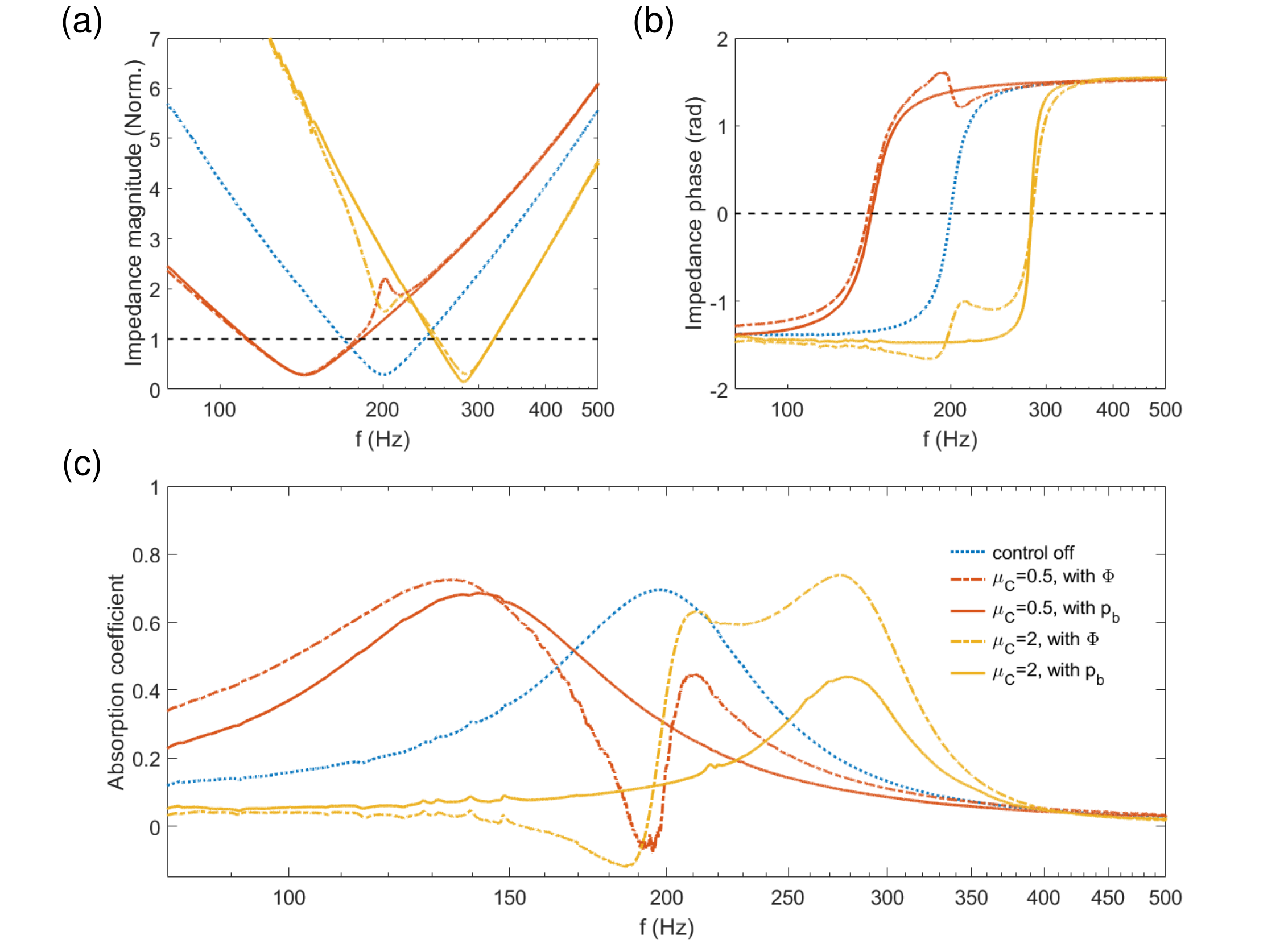}
	\centering
	\caption{\label{fig: Control_compare} The measured specific acoustic impedance (magnitude normalized by $Z_c$ (a), phase (b)) and the absorption coefficient (c) of the achieved AER, subjected to the former FF-AER control (dash-dotted line curves) and the currently proposed PID-AER control (plain line curves) respectively. The compliance adjustment with laws of $\mu_C=0.5$ and $2$ are considered in both control schemes (the resistance and the mass of the controlled ER are maintained with $\mu_M=\mu_R=1$). The legend in (c) is used for all three figures.}
\end{figure}

We have already compared in the analytical study of section \ref{Analytical} the currently proposed PID-AER method with the one formerly reported by Rivet et al. \cite{etienne2016}. In this section, the comparison will be held experimentally instead. Taking the example of compliance adjustments with $\mu_C=0.5$ and $2$ respectively, Fig.~\ref{fig: Control_compare} shows the corresponding measured impedance (magnitude and phase) and absorption coefficient curves of the two control techniques, namely the model-based FF-AER strategy \cite{etienne2016} and the proposed PID-AER method. 

In the FF-AER control approach, the influence of time delay manifests especially around the natural resonance of the passive ER. It can thus lead to a negative absorption coefficient, as evidenced by the experimental results reported on Fig.~\ref{fig: Control_compare} (yellow and red dash-dotted line curves) which remains consistent with the analytical study of section \ref{Analytical}. Meanwhile, the inaccuracy occurring in the parameter estimation (e.g., loudspeaker's Thiele-Small parameters, enclosure volume, etc.) can also affect the control results. Since these discrepancies affecting the control accuracy are impossible to compensate in a straightforward manner, a more elaborate control architecture should be implemented. 

\begin{figure}[htbp]
	\includegraphics[width=0.75\textwidth]{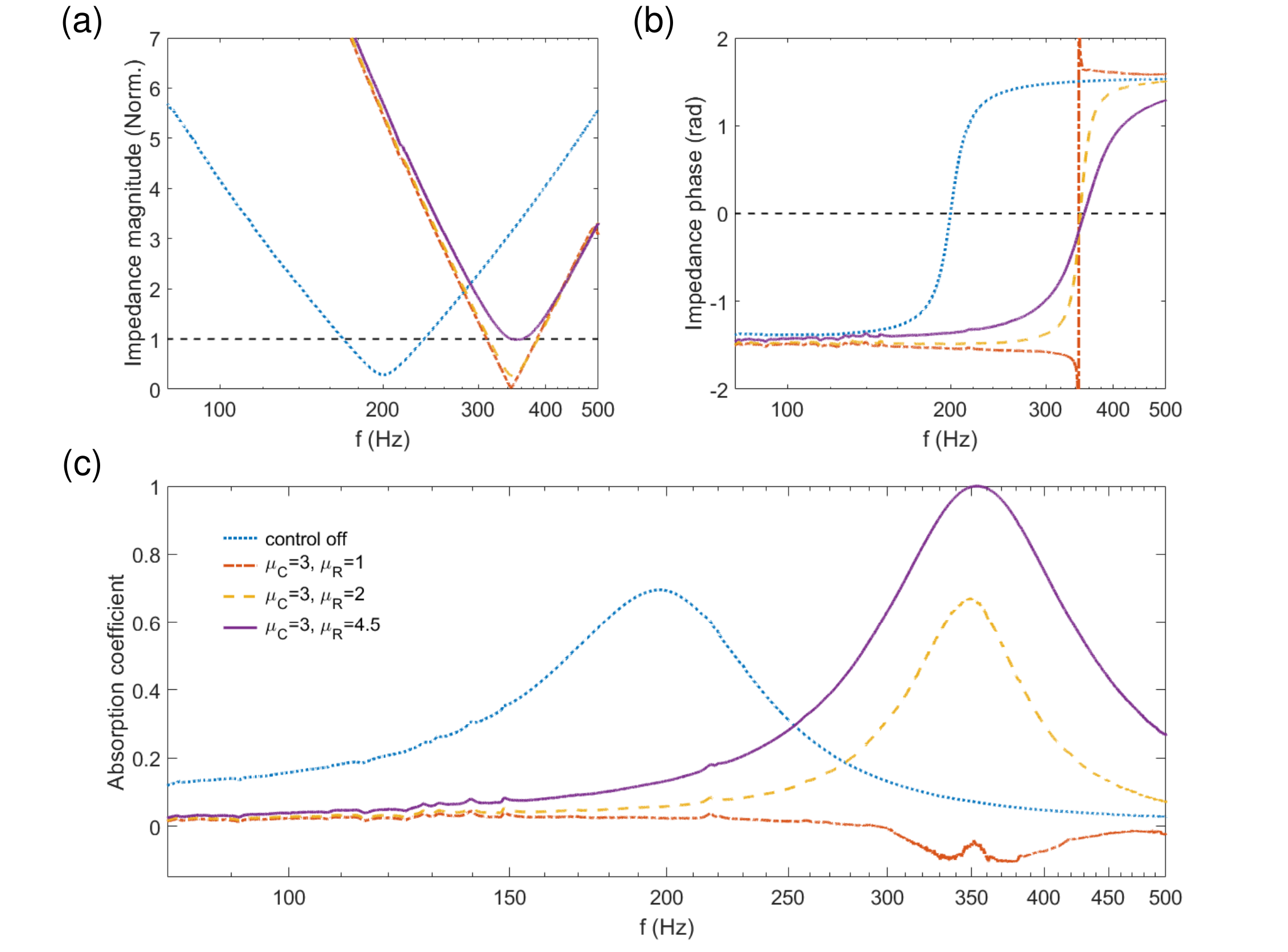}
	\centering
	\caption{\label{fig: Control_limit} The measured specific acoustic impedance (magnitude normalized by $Z_c$ (a), phase (b)) and the absorption coefficient (c) of the achieved AER, subjected to the proposed PID-AER control with a law of $\mu_C=3$ and $\mu_M=1$. The resistance of the controlled ER is tuned from $\mu_R=1$ to $\mu_R=2$ and then $\mu_R=3.5$, to allow finally a perfect absorption to be achieved at the target resonance frequency. The legend in (c) is used for all three figures.}
\end{figure}\textbf{}

On the contrary, in the proposed PID-AER control method, the time delay has mainly a marginal impact on the magnitude around the target resonance (see plain line curves in Fig.~\ref{fig: Control_compare}). In the experimental studies of section \ref{AER_results}, it has been already demonstrated that this effect can be easily compensated by a change in the P gain (acting on resistance) through assigned parameter $\mu_R$. Compared to the FF-AER method, the proposed PID-AER control allows the acoustic performance of the AER to be modified without any mismatch around the natural resonance, it provides more accuracy in the whole frequency range, enabling the achievement of a SDOF resonator with resonance behavior perfectly in line with the expectations.

When the compliance/mass control is considered to move the resonance of the AER farther from its natural resonance, a more significant influence of the time delay is observed on the target resonance magnitude, as presented in section \ref{AER_results}. In Fig.~\ref{fig: Control_limit}, a limit case of the compliance adjustment implemented with a law of $\mu_C=3$ is illustrated, which causes the resonance of the AER to shift from $\SI{200}{}Hz$ to $\SI{355}{Hz}$. In this case, the time delay leads to a negative magnitude of absorption coefficient around the target resonance, meaning that the AER will re-inject energy into the external environment, exactly as predicted in the analytical study of section \ref{Analytical}. Such unstable state can result in the production of a whistling at the resonance modes of the tube in front of the loudspeaker. Whereas, by performing the additional resistance adjustment, the AER can be re-stabilized to absorb energy. The yellow color curves in Fig.~\ref{fig: Control_limit} depict the control results when the parameter $\mu_R$ is set to $2$ together with $\mu_C=3$, the absorption coefficient of the AER is increased from negative value to positive ones, to reach the same level as the passive case (blue dotted line curves). Then, if we continue to increase the value of parameter $\mu_R$ to $4.5$, a perfect absorption (with $\alpha=1$) can finally be attained at the target resonance frequency. 

Therefore, based on all the above assessments, the proposed PID-AER control approach shows its ability to precisely redefine the acoustic properties of the controlled resonator. The tunable frequency range of the achieved resonance can be further expanded through a judicious choice of the control law parameters.

\section{Conclusion}
Considering an ER made of a closed-box electrodynamic loudspeaker as the basic actuator, a current-driven PID-like control has been proposed and explored in the present work. Thanks to the proportionality between the pressure inside the enclosure and the axial displacement of the membrane, valid within the frequency range of interest ($[\SI{50}{Hz}, \SI{500}{Hz}]$), such control approach was based on the estimation of the membrane axial velocity, displacement and the overall net forces acting on the membrane. It has been implemented through applying three individual feedback gains on these dynamic quantities derived from the pressures inside the enclosure and in front of the membrane sensed in real time, enabling the adjustment of the resistance (by a P-like method), the compliance (by a I-like method) and the moving mass (by an equivalent D-like method) of the AER, respectively.

Starting with an analytical study, a preliminary evaluation on the feasibility and efficiency of the control has been carried out, where the time delay inherent to the control execution was accounted for. Thereafter, the research has focused on experimental realization of an AER according to the reported PID-like control scheme. It has been demonstrated that each of the resonator characteristics (compliance, resistance and moving mass) can be individually tuned in a prescribed manner to allow the resonance behavior of the AER, namely its resonance frequency and bandwidth, to be tailored as expected. A good agreement has been found between the experimental results and the analytical predictions, ensuring that the control was operated as defined.

With a view to improving the sound absorption, the resistance adjustment has been accounted for either individually, or to combine with the compliance or/and the mass adjustments. In each type of control combination, the absorption property of the AER can always be optimized to the greatest extent, enabling the absorption bandwidth to be enlarged and the perfect absorption (absorption coefficient equal to $1$) to be reached at the desired frequency. When the mass and the compliance of the AER are both scaled by $0.4$ through the control, tuning simultaneously the resistance has lead to an effective absorption (with absorption coefficient greater than $0.83$) in the frequency range of $[\SI{130}{Hz},\SI{300}{Hz}]$. Compared with the feedforward impedance control \cite{etienne2016}, the proposed PID-AER control has proven both numerically and experimentally to be more accurate. It allows an easier compensation of both the mismatch introduced by the control time delay and the discrepancy existed in the parameter estimations (required for defining the control laws), through directly a tuning of control parameter settings. 

In the current study, in order to ensure the linear relation between the pressure inside the enclosure and the membrane axial displacement, we have limited the frequency range to below $\SI{500}{Hz}$. Indeed, the possible operating frequency range for implementing the proposed PID-like control is related to the dimension and the resonance frequency of the resonator used. For instance, if we consider a loudspeaker of model Visaton FRWS 5-8 Ohms with an enclosure of $\SI{9}{cm}\times\SI{9}{cm}\times\SI{5}{cm}$ which has a resonance frequency of around $\SI{500}{Hz}$, the frequency range of interest can be extended up to $\SI{1.2}{kHz}$ to perform the PID-AER control. 

For the future, the developed PID-AER control strategy can be used to combine with other types of linear or nonlinear controls (such as the one presented in \cite{Xinxin_PRApplied}) to further develop a hybrid control with promising accuracy and/or stability. The digital implementation of control through FPGA platform offers more flexibility and convenience for defining and evaluating different control laws, however it becomes costly when many resonators need to be controlled. As a perspective, it can serve as a guide for other low-cost solutions. For instance, an analogous impedance control can be performed by considering the same control laws as reported here, but by shunting the ER terminals with an electrical network which can be identified within the proposed PID-like control. In addition, according to the capacity of the control platform we used, our studies could also be devoted to designing an acoustic liner or metasurface with limited number of active unit cells to enable either a broadband absorption or other specific wave manipulations. 

\section{acknowledgments}
This work was supported by the Swiss National Science Foundation (SNSF) under grant No. ${200020}\_{200498}$.





\end{document}